\documentclass[11pt]{article}
\usepackage{amsmath,amsfonts,setspace,cancel,url,hyperref,verbatim,slashed,amsthm,graphicx,color,cite,todonotes,tikz-cd}
\usepackage{xcolor}
\usepackage{hyperref}
\usepackage{dsfont}
\usepackage[T1]{fontenc}
\usepackage[utf8]{inputenc}
\usepackage{mathtools}

\usepackage{fullpage}
\newcommand{\be}{\begin{equation}}
\newcommand{\ee}{\end{equation}}

\newcommand{\gM}{\mathcal{M}}
\newcommand{\cH}{\mathcal{H}}
\newcommand{\hmu}{\hat{\mu}}
\newcommand{\hnu}{\hat{\nu}} 
\newcommand{\hrho}{\hat{\rho}}

\newcommand{\hsigma}{\hat{\sigma}}

\newcommand{\ba}{\bar{a}}
\newcommand{\bb}{\bar{b}}
\newcommand{\bh}{\bar{h}}

\newcommand{\bk}{\bar{k}}

\newcommand{\bm}{\bar{m}}
\newcommand{\bn}{\bar{n}}
\newcommand{\bX}{\bar{X}}
\newcommand{\bY}{\bar{Y}}

\newcommand{\hv}{\hat{v}}

\newcommand{\Aa}{\mathcal{A}}
\newcommand{\Ab}{\mathcal{B}}
\newcommand{\Ac}{\mathcal{C}}

\newcommand{\Gfour}{\mathrm{SL}(5)}
\newcommand{\Gfive}{\mathrm{SO}(5,5)}
\newcommand{\Gsix}{E_{6(6)}}
\newcommand{\Gseven}{E_{7(7)}}
\newcommand{\Geight}{E_{8(8)}}
\newcommand{\Edd}{E_{d(d)}}

\newcommand{\Hfour}{\mathrm{SO}(5)}
\newcommand{\Hfive}{\mathrm{SO}(5)\times \mathrm{SO}(5)}
\newcommand{\Hsix}{\mathrm{USp}(8)}
\newcommand{\Hseven}{\mathrm{SU}(8)}
\newcommand{\Height}{\mathrm{SO}(16)}

\newcommand{\HfourLor}{\mathrm{SO}(2,3)}
\newcommand{\HfiveLor}{\mathrm{SO}(5, \mathbb{C})}
\newcommand{\HsixLor}{\mathrm{USp}(4,4)}
\newcommand{\HsevenLor}{\mathrm{SU}^*(8)}
\newcommand{\HeightLor}{\mathrm{SO}^*(16)}

\newcommand{\gendil}{\mathbf{d}}

\newcommand{\VR}{V}
\newcommand{\VL}{\bar{V}}

\newcommand{\bA}{\bar{A}}
\newcommand{\bB}{\bar{B}}

\newcommand{\bmu}{\bar{m}}

\numberwithin{equation}{section}

\begin{document}


\title{
\begin{flushright}
\normalsize{QMUL-PH-19-04}
\end{flushright}
\bigskip
\bf
Non-Riemannian geometry of M-theory 
}

\author{\sc David S. Berman${}^1$\footnote{\tt d.s.berman@qmul.ac.uk}, Chris D. A. Blair${}^2$\footnote{\tt cblair@vub.ac.be}{}\, and Ray Otsuki${}^1$\footnote{\tt r.otsuki@qmul.ac.uk}}

\date{\it $^1$ Centre for Research in String Theory, School of Physics and Astronomy,
Queen Mary University of London, 327 Mile End Road, London E1 4NS, UK 
\\ 
$^2$ Theoretische Natuurkunde, Vrije Universiteit Brussel, and the International Solvay Institutes,  Pleinlaan 2, B-1050 Brussels, Belgium 
}

\maketitle

\begin{abstract}
We construct a background for M-theory that is moduli free. 
This background is then shown to be related to a topological phase of the $\mathrm{E}_{8(8)}$ exceptional field theory (ExFT). 
The key ingredient in the construction is the embedding of non-Riemannian geometry in ExFT.
This allows one to describe non-relativistic geometries, such as Newton-Cartan or Gomis-Ooguri-type limits, using the ExFT framework originally developed to describe maximal supergravity. 
This generalises previous work by Morand and Park in the context of double field theory.

\end{abstract}

\tableofcontents

\section{Introduction}

General relativity describes the geometry of gravity in terms of a dynamical (pseudo-)Riemannian metric. 
String theory and M-theory provide a route towards a quantum mechanical understanding of gravity. 
At low energies, the classical geometry of string theory/M-theory is again described by a metric, whose dynamics is governed by a supergravity theory in which the metric is accompanied by a collection of scalars and $p$-form gauge fields, plus fermions.

The presence of duality in these theories means that they exhibit a (hidden) symmetry which mixes metric and form field components.
Inspired in large part by a desire to capture and explain this symmetry more fundamentally, and to find new notions of intrinsically ``stringy'' or ``M-theoretic''  geometry treating all the massless states of the theory on a more egalitarian footing, reformulations of the dynamics of supergravity have been found in which the geometry and the fields living in the geometry are united and covariance under the duality groups of string/M-theory is made manifest. These efforts have led to the modern development of double field theory (DFT) and exceptional field theory (ExFT) \cite{Hull:2009mi,Hillmann:2009ci, Berman:2010is,Berman:2011cg, Berman:2011jh, Berman:2012vc, Hohm:2013vpa, Hohm:2013uia, Hohm:2014fxa}, building on pioneering earlier work such as \cite{Duff:1989tf,Tseytlin:1990nb, Siegel:1993th, Siegel:1993xq, West:2001as, West:2003fc} and on the introduction of generalised geometry \cite{Gualtieri:2003dx, Hitchin:2004ut}.

The starting point for these theories is to observe that the bosonic degrees of freedom of supergravity in a certain $(n+d)$-dimensional split can be recombined into multiplets of the groups $\mathrm{O}(d,d)$ (when $n+d=10$ \cite{Hohm:2013nja} with the original construction applicable to $n=0$ \cite{Hull:2009mi,Hohm:2010pp}) or $\Edd$ (when $n+d = 11$, and so far allowing for $d=2,\dots,9$ \cite{Hohm:2013vpa, Hohm:2013uia, Hohm:2014fxa, Hohm:2015xna, Abzalov:2015ega, Musaev:2015ces, Berman:2015rcc, Bossard:2018utw}). 
Then the full dynamics and local symmetries of 10- or 11-dimensional supergravity can be encoded in a formulation with a manifest covariance under $\mathrm{O}(d,d)$ or $\Edd$.
The usual diffeomorphism symmetry, which is associated to the group $\mathrm{GL}(d)$, is extended to a notion of generalised diffeomorphisms involving local $G$ ($=\mathrm{O}(d,d)$ or $\Edd$) transformations and realised using an extended set of coordinates $Y^M$ transforming under a particular representation of $G$. The original theories are recovered by solving a constraint known as the ``section condition'' which restricts the dependence of all fields in the theory to a subset of the $Y^M$.  

Different solutions of the section condition lead to different parametrisations of the fundamental DFT/ExFT variables in terms of standard supergravity fields, depending on different choices of the physical coordinates amongst the $Y^M$. In this way, for instance, ExFT admits inequivalent solutions of the section condition giving either 11-dimensional supergravity or the 10-dimensional type IIB supergravity \cite{Hohm:2013vpa, Blair:2013gqa}. One can think of the usual supergravity theories as following from the single unifying ExFT formulation on solving the consistency conditions of the latter. 

A more ambitious interpretation of the geometry of DFT/ExFT is to allow for solutions of the section condition, or parametrisations of the fields, which \emph{do not reduce to conventional supergravity}. A number of avenues have been explored, often involving notions of ``non-geometry'' in one form or another (for lots on non-geometry, see the review \cite{Plauschinn:2018wbo}).
This includes the possibility of relaxing the section condition in order to carry out Scherk-Schwarz type reductions where the twist matrices may depend on dual coordinates.
This leads to lower-dimensional gauged supergravities including those that are not possible to lift to ten- or eleven-dimensional geometries \cite{Dibitetto:2012rk} as well as, in ten dimensions, the massive IIA theory \cite{Hohm:2011cp, Ciceri:2016dmd} and the so-called ``generalised supergravity''. The latter was remarkably only discovered as a type IIB string background less than three years ago \cite{Wulff:2016tju} (from examining in close detail the relationship between kappa symmetry and the string beta function) and was almost immediately shown to have a simple DFT and ExFT picture \cite{Sakatani:2016fvh, Baguet:2016prz}. Furthermore, the DFT/ExFT geometry also provides a home for explicit solutions corresponding to strongly non-perturbative states known as exotic branes -- whose existence is predicted by U-duality \cite{Obers:1998fb,deBoer:2012ma} -- that have no global description in supergravity \cite{Blair:2014zba, Bakhmatov:2017les,Fernandez-Melgarejo:2018yxq,Berman:2018okd}.
In some instances it is possible to describe non-geometry by using alternative spacetime parametrisations of the DFT/ExFT fields, for instance in terms of a metric and bivector \cite{Grana:2008yw,Andriot:2012an}, rather than a metric and two-form. 

The examples of the previous paragraph are still based on the idea that there is some spacetime description involving (possibly only locally) a Riemannian metric and some set of forms or bivectors. The novelty in the exotic backgrounds arises from global data as we ``glue'' patches using $E_{d(d)}$ transformations, rather than traditional diffeomorphisms, but locally there is a supergravity description of some sort though perhaps gauged or ``generalised'' due to a Scherk-Schwarz twist. These non-geometric aspects of DFT/ExFT are of course of crucial importance. There is, however, a further generalisation we can make that will be the subject of this paper.

This is the, perhaps rather unexpected, observation that DFT/ExFT also accommodates descriptions of non-Riemannian geometry. These are backgrounds where there is not an invertible spacetime metric but instead a non-relativistic geometry, or even no intrinsic geometric structure at all. Examples of such geometries go back to Newton-Cartan geometry \cite{Cartan:1923zea, Cartan:1924yea}, and include the non-relativistic limit of string theory studied by Gomis and Ooguri \cite{Gomis:2000bd} and Danielsson, Guijosa and Kruczenski \cite{Danielsson:2000gi,  Danielsson:2000mu}.

The exploration of non-Riemannian string theory geometries using DFT was pioneered in \cite{Lee:2013hma,Ko:2015rha,Morand:2017fnv,Cho:2018alk}, leading to a classification of allowed non-Riemannian backgrounds \cite{Morand:2017fnv}. The starting point of these papers was the realisation that backgrounds defined by a generalised metric without a conventional spacetime interpretation \cite{Lee:2013hma} could still be studied using the doubled sigma model \cite{Tseytlin:1990nb, Hull:2004in}, which describes a string whose target space is the doubled geometry of DFT. This was connected to the non-relativistic Gomis-Ooguri string in \cite{Ko:2015rha}, including an analysis of the spectrum using spacetime DFT techniques. The paper \cite{Morand:2017fnv} then offered a \emph{classification} of non-Riemannian parametrisations of $\mathrm{O}(d,d)$ generalised metrics, as well as a study of particle and string actions in these backgrounds. 

In this paper, we will study examples of M-theory non-Riemannian geometry in the context of exceptional field theory,\footnote{Previously in \cite{Park:2014una} there was a limited discussion of some example non-Riemannian parametrisation in a class of theories generalising the internal part of the $\Gfour$ ExFT to $\mathrm{SL}(N)$.} showing that they provide connections to non-relativistic and topological theories within a framework originally intended to describe maximal supergravity. 

\subsubsection*{Non-Riemannian backgrounds in $\mathrm{O}(d,d)$: a first encounter}

To set the scene, let us give a brief review of the ideas appearing in the non-Riemannian $\mathrm{O}(d,d)$ backgrounds studied in \cite{Lee:2013hma,Ko:2015rha,Morand:2017fnv,Cho:2018alk}.
We will focus on ``generalised metric'', which may be interpreted as describing the geometry of the extended spacetime with coordinates $Y^M$, with $M=1,\dots, 2d$, in the fundamental representation of $\mathrm{O}(d,d)$.
We can define the generalised metric solely by the properties of being symmetric and compatible with the $\mathrm{O}(d,d)$ structure $\eta$.
Hence, denoting it by $\cH_{MN}$, it obeys:
\be
\cH \eta^{-1} \cH = \eta \, ,  \label{comptcon}
\ee where 
\be
\eta_{MN}  = \begin{pmatrix} 0 & I \\ I & 0 \end{pmatrix} \,.
\ee
In the usual DFT formulation, one then solves the compatibility condition (\ref{comptcon}) by parametrising $\cH$ as follows:
\be
\cH_{MN} = \cH_{MN}(g,B_2) = 
\begin{pmatrix} g - B g^{-1} B & B g^{-1} \\ - g^{-1} B & g^{-1} \end{pmatrix} \,,
\label{Hnormal}
\ee
so that subsequently it is possible to interpret $g$ as the spacetime metric and $B_2$ the Kalb-Ramond two-form of a $d$ dimensional space. 
This generalised metric is also encountered in the usual string worldsheet theory, where it appears in the Hamiltonian form of the string worldsheet action:
\be
S = \int d^2\sigma \dot{X}^i P_i - \frac{1}{2} Z^M \cH_{MN} Z^N 
\,,\quad
Z^M \equiv \begin{pmatrix} {X}^{\prime i} \\ P_i \end{pmatrix} \,.
\ee
Here, if we integrate out the momenta $P_i$ we recover the usual Polyakov string action in conformal gauge, with background metric and $B$-field. 
This Hamiltonian action is actually very closely related to the doubled approach to the string sigma model, in which the target space has doubled coordinates $X^M = ( X^i, \tilde X_i)$ and $\tilde X_i^\prime = P_i$. 
However, to illustrate how the generalised metric describes ``non-Riemannian'' backgrounds, we will continue here to think in terms of the Hamiltonian of the usual string, rather than in DFT. 

One of the advantages of the Hamiltonian picture is that it frequently allows one to take limits which would be singular in the Lagrangian formulation.
Here, this manifests itself in the ability to choose $\cH_{MN}$ to have a degenerate bottom right $d \times d$ block. This is the block that would normally be interpreted as describing the inverse spacetime metric. However, the non-degeneracy of this block is compensated within $\cH_{MN}$ such that the whole generalised metric remains well-defined.
An example of such a situation in the case $d=2$ (suppressing the other target space coordinates) is given by
\be
\cH_{MN} = \begin{pmatrix} 2 \mu \eta_{ij} & Z_i{}^j \\ Z_j{}^i & 0 \end{pmatrix} 
\,,\quad
\eta_{ij} = \text{diag}\,(-1,1)
\,,\quad
Z_i{}^j = \begin{pmatrix} 0 &1 \\ 1 & 0 \end{pmatrix}\,.
\ee
The momenta $P_i$ cannot be integrated out but, rather, act to impose chirality conditions:
\be
S = \int d^2 \sigma P_i ( \dot{X}^i - Z_j{}^i X^{\prime j} ) - \mu \eta_{ij} X^{\prime i} X^{\prime j} \,.
\ee
The generalised metric is that which you obtain in the non-relativistic limit of string theory due to Gomis and Ooguri \cite{Gomis:2000bd}.

A more striking example is to set $\cH_{MN} = \eta_{MN}$ (for any $d$). This choice clearly solves (\ref{comptcon}) but is not expressible as parameterised in (\ref{Hnormal}). 
In the classification of \cite{Morand:2017fnv}, this has the special position of being ``maximally non-Riemannian''.
 The resulting sigma model is described by
\be
S = \int d^2 \sigma P_i ( \dot{X}^i - X^{\prime i} ) 
\ee
an entirely chiral theory, which appears to be related to recent work on certain chiral or ``twisted'' string theories with unusual properties \cite{Siegel:2015axg,Casali:2016atr, Casali:2017mss,Lee:2017utr,Lee:2017crr} or a beta-gamma system \cite{Nekrasov:2005wg}. The full quantum consistency of this model and its possible supersymmetrisation is a fascinating topic for future work.
Viewed as an admissible background in DFT, this case has some very interesting properties: for instance, arbitrary variations $\delta \cH_{MN}$ evaluated on this background are automatically projected to zero \cite{Cho:2018alk}. Thus the maximally non-Riemannian space has no moduli!

\subsubsection*{Outline of this paper}

The goal of this paper is to study the M-theory theory analogues of such backgrounds via exceptional field theory. 
This will involve re-analysing the consistency conditions on the generalised metrics of ExFT. 
Although in the string theory case, we have access to sigma models (either the Hamiltonian form presented above or the truly doubled sigma models such as \cite{Tseytlin:1990nb, Hull:2004in}) with which to explore the non-Riemannian background, the nature of $\Edd$ covariant worldvolume theories for M-branes (if such theories exist) is mysterious \cite{Duff:2015jka} (though see \cite{Arvanitakis:2017hwb,Arvanitakis:2018hfn, Sakatani:2017vbd} for some ExFT inspired approaches). 
Nevertheless, some of the first examples of $\Edd$ generalised metrics (for $d=4$, i.e. $\Gfour$) were found by studying the M2 worldvolume theory \cite{Duff:1990hn, Berman:2010is} (in fact they appear also in the M2 Hamiltonian). We therefore expect, or hope, that the geometry of ExFT more generally tells us something about the structure of M-theory backgrounds beyond the conventional geometry. 

We will also focus on the generalisation of the DFT maximally non-Riemannian background $\cH_{MN} = \eta_{MN}$. This depends on the presence of an invariant tensor in the symmetric product of $R_1 \otimes R_1$.
For the exceptional generalised diffeomorphism groups $\Edd$, the representation theory precludes the existence of such a tensor in all finite dimensional cases except that of $\Geight$. 
In this case, $R_1 = \mathbf{248}$ is the adjoint (also the fundamental), and we therefore can define $\gM_{MN}$ to be proportional to the Killing form $\kappa_{MN}$.
We propose to view the $\Geight$ ExFT on the background $\gM_{MN} = - \kappa_{MN}$ as the definition of the topological ``phase'' introduced in \cite{Hohm:2018ybo} and used to study three-dimensional superconformal field theories \cite{Hohm:2018git}. 
In that work, an \emph{ad hoc} truncation was taken by setting $\gM_{MN} = 0$ in order to preserve the full $\Geight$ symmetry. By extending the set of allowed backgrounds to include non-Riemannian generalised metrics, we obtain a non-singular definition of the topological theory.
It is tempting to speculate that this is connected to an old idea (touched upon in for instance \cite{Tseytlin:1981ks, Witten:1988xj, Horowitz:1989ng}) that there should be some underlying topological phase of gravity and that the geometry may emerge through spontaneous breaking. What is interesting about the proposal using DFT and ExFT is that the topological phase exists not with a vanishing metric, as originally envisaged, but via a moduli-free maximally non-Riemannian metric.

This paper is structured as follows. 
After the description of the main result of the topological $E_8$ vacuum in section two, we discuss the non-Riemannian backgrounds in $O(d,d)$ DFT and the relationship to Newton-Cartan geometry and the Gomis-Ooguri geometry for the closed string. 
Our intention is to lift these ideas to ExFT. Hence we first in section four introduce the $\Gfour$ ExFT and its various parameterisations. 
This allows us in section five to extend what was done in terms of DFT to the $\Gfour$ theory. 
In particular, we first play with some simple examples of ExFT backgrounds which lead to non-Riemannian parametrisations, including the Gomis-Ooguri limit for M2 branes, the (timelike) U-dual of the M2 brane solution, and related examples on the IIB side. 
We then establish a general parametrisation of the generalised metric of the $\Gfour$ theory which can be used to describe such examples. We discuss the symmetries of this parametrisation, including some ambiguities involving shift symmetries that are inherent to the non-Riemannian parametrisation. We further relate our parametrisation to the DFT case by reducing from $\Gfour$ to $O(3,3)$. 
Finally, we offer a number of ways to embed (M-theoretic versions of) Newton-Cartan non-relativistic geometry in the $\Gfour$ ExFT.

Our intention is thus to demonstrate the utility of the ExFT formulation for describing backgrounds of more general theories than the standard supergravities for which the theory was initially introduced. 
This leads to non-relativistic theories of gravity, topological three-dimensional theories, and more. We hope that this paper will stimulate further interest in these uses of DFT/ExFT and provide a starting point to study features of theories with non-Riemannian or non-relativistic geometries. 

\section{Generalised metrics, projectors and the topological $\Geight$ vacuum}

\subsection{Generalised metrics and diffeomorphisms}

The local symmetries of general relativity, double field theory and exceptional field theory can all be treated in same manner, by defining (generalised) diffeomorphisms associated to a group $G$. 
For general relativity, this group is $G = \mathrm{GL}(d)$, for DFT, it is $G = \mathrm{O}(d,d)$, and for ExFT, it is $\Edd$. 
We work with coordinates $(X^\mu, Y^M)$, where $\mu = 1,\dots,n$ and $Y^M$ transform in what we call the $R_1$ representation of $G$. In DFT and ExFT, we will call the $X^\mu$ coordinates ``external'' and the $Y^M$ ``internal'' or ``extended'', mimicking the language we would use if we reduced to an $n$-dimensional theory (however no compactification is assumed or needed to formulate these theories).
The $R_1$ representation is the $d$-dimensional fundamental of $\mathrm{GL}(d)$ in the case of general relativity, the $2d$-dimensional fundamental in the case of $\mathrm{O}(d,d)$, and for $\Edd$ the representations are listed in table \ref{GHR} (the rule is that $R_1$ is the representation whose highest weight is the fundamental weight associated to the rightmost node on the Dynkin diagram).

We define (generalised) diffeomorphisms associated to the transformation of the coordinates $\delta Y^M = - \Lambda^M$ in terms of a (generalised) Lie derivative acting on vectors $\delta_\Lambda V^M = \mathcal{L}_\Lambda V^M$ by
\be
\mathcal{L}_\Lambda V^M = \Lambda^N \partial_N V^M - \alpha \mathbb{P}_{adj}{}^M{}_K{}^N{}_L \partial_N \Lambda^L V^K + \lambda_V \partial_K \Lambda^K V^M\,,
\label{gldproj}
\ee
where $\mathbb{P}_{adj}{}^M{}_K{}^N{}_L$ denotes the projector from $R_1 \otimes \bar{R}_1$ onto the adjoint representation, $\alpha$ is a constant which depends on the group under consideration (see table \ref{GHR}) and $\lambda_V$ denotes the weight of $V^M$. 
It is often useful to expand the projector to obtain an equivalent form of the generalised Lie derivative:
\be
\mathcal{L}_\Lambda V^M = \Lambda^N \partial_N V^M -  V^N \partial_N \Lambda^M   + Y^{MN}{}_{KL} \partial_N \Lambda^K V^L + ( \lambda_V + \omega)  \partial_K \Lambda^K V^M \,,
\label{gldY}
\ee
which makes apparent how the structure differs from the ordinary Lie derivative (which is given by the first two terms). 
The modification involves the so-called $Y$-tensor, which is constructed out of group invariants \cite{Berman:2012vc} (for instance, for $\mathrm{O}(d,d)$, $Y^{MN}{}_{KL} =\eta^{MN} \eta_{KL}$), and also a constant $\omega$ which can be thought of as an intrinsic weight. When $G = \mathrm{GL}(d)$, clearly $Y^{MN}{}_{KL} = 0$ and $\omega = 0$.

\begin{table}\centering
\begin{tabular}{ccccccc} 
$G$ & $H$ & $H^*$ & $\alpha$ & $\omega$ & $R_1$ & $R_2$ \\\hline 
$\mathrm{GL}(d)$ &  $\mathrm{SO}(d)$ & $\mathrm{SO}(1,d-1)$ & 1 & 0 & $\mathbf{d}$ & n/a \\
$\mathrm{O}(D,D)$ & $\mathrm{O}(D) \times \mathrm{O}(D)$ & $\mathrm{O}(1,D-1) \times \mathrm{O}(1,D-1)$ & $2$ & $0$ & $\mathbf{2D}$ & $\mathbf{1}$ \\
$\Gfour$ & $\Hfour$ & $\HfourLor$ & $3$ & $-1/5$ & $\mathbf{10}$ & $\mathbf{\bar{5}}$ \\
$\Gfive$ & $\Hfive$ & $\HfiveLor$& $4$ & $-1/4$ & $\mathbf{16}$ & $\mathbf{10}$\\
$\Gsix$ & $\Hsix$ & $\HsixLor$ & $6$ & $-1/3$ & $\mathbf{27}$ & $\mathbf{\bar{27}}$\\
$\Gseven$ & $\Hseven$ & $\HsevenLor$& $12$  &$-1/2$ &$\mathbf{56}$ & $\mathbf{133}$\\
$\Geight$ & $\Height$ & $\HeightLor$& $60$ & $-1$ & $\mathbf{248}$ & $\mathbf{1 \oplus 3875}$\\
\end{tabular}
\caption{The vital statistics of ordinary geometry, DFT and ExFT. The (generalised) metric lives in $G/H$ (Euclidean case) or $G/H^*$ (Lorentzian case) \cite{Hull:1998br}, (generalised) vectors are valued in $R_1$, and the section condition in $R_2$. The intrinsic weight is given by $\omega = -1/(n-2)$ in ExFT and $\omega = 0$ in DFT.}
\label{GHR} 
\end{table} 

The crucial distinction between the $\mathrm{GL}(d)$ Lie derivative of usual Riemannian geometry and the $\mathrm{O}(d,d)$ or $\Edd$ generalised Lie derivative is that though the former leads to a closed symmetry algebra (closed under the Lie bracket), the algebra of generalised Lie derivatives turns out to be obstructed. The root cause of this obstruction is the dependence of the fields and gauge parameters on the coordinates $Y^M$. One way to guarantee closure is then to impose the \emph{section condition} on the coordinate dependence of all fields and gauge parameters, which can be realised as the condition that $\partial_M \otimes \partial_N \big|_{R_2} = 0$ or (in most cases) simply that $Y^{MN}{}_{KL} \partial_M \otimes \partial_N = 0$. Here the derivatives may act on separate quantities or on a single quantity.  
Solutions to the section condition will break the (global and local) $G$-symmetry and amount to a choice of $d$ coordinates $Y^i$ from amongst the $Y^M$ which are taken to be physical. This is how one reduces the formulation with manifest $G$-covariance to the standard geometric description. When further isometries are present such that $\partial_M =0$ then the global group $G$ remains a symmetry and is identified with a usual duality group. Essentially, duality arises from the ambiguity in identifying the physical spacetime inside the extended space when there are isometries. 

The geometry of general relativity is, of course, described by a metric. Similarly the generalised, or ``extended'', geometry of DFT/ExFT will be described by a generalised metric. 
We define this to be a symmetric matrix, $\gM_{MN}$, which is an element of $G$ and so preserves the appropriate invariant tensors. The generalised Lie derivative of the generalised metric follows from \eqref{gldproj} or \eqref{gldY} using the Leibniz property. It takes the form:
\be
\delta_\Lambda \gM_{MN} = \Lambda^P \partial_P \gM_{MN} + 2 \alpha P_{MN}{}^{KL} \partial_K \Lambda^P \gM_{LP} \,,
\label{lgm}
\ee
in which the following projector appears:
\be
P_{MN}{}^{KL} = \frac{1}{\alpha} \left( \delta^{(K}_M \delta^{L)}_N - \omega \gM_{MN} \gM^{KL} - \gM_{MQ} Y^{Q(K}{}_{RN} \gM^{L) R}\right) \, ,
\label{cosetPY}
\ee
or in terms of the adjoint projector,
\be
P_{MN}{}^{KL} = \gM_{MQ} \mathbb{P}_{adj}{}^Q{}_N{}^{(K}{}_R \gM^{L) R} \,.
\label{cosetPP}
\ee
Note that as the $Y$-tensor, or equivalently the adjoint projector, is a group invariant it is preserved by the simultaneous action of $\gM$ and $\gM^{-1}$ on all four indices, which can be used to check that $P_{MN}{}^{KL}$ is actually symmetric in both its upper and lower pairs of indices.
We can think of equation \eqref{lgm} as expressing the variation of the generalised metric, in terms of a parameter $\partial_{(K} \Lambda^P \gM_{L) P}$, which is then projected from the symmetric tensor product of $R_1$ with itself into the space in which $\gM_{MN}$ lives by means of $P_{MN}{}^{KL}$. Generically, $\gM_{MN}$ is in fact valued in a coset $G/H$. 

We can calculate the trace of the projector to compute the number of independent components of the generalised metric, i.e. the dimension of the coset $G/H$ in which it lives. 
In general, we find:
\be
P_{MN}{}^{MN} = \frac{1}{2 \alpha} \left(
\mathrm{dim} R_1 ( \mathrm{dim} R_1 + 1 - 2 \omega ) - Y^{MN}{}_{MN} - \gM_{MN} Y^{MN}{}_{PQ} \gM^{PQ} 
\right)\,.
\label{traceP}
\ee
Evidently, in general relativity we have $\alpha = 1$, and the terms in \eqref{cosetPY} involving $\omega$ and the $Y$-tensor do not appear. Hence we find $P_{MN}{}^{MN} = \frac{1}{2} d (d+1)$ which is the number of independent components of a symmetric matrix and also the dimension of the coset $\mathrm{GL}(d)/\mathrm{SO}(d)$. 

In DFT and ExFT the situation is rather more interesting.
Part of the trace \eqref{traceP} is independent of the generalised metric and follows from representation theory as the Y-tensor can be related to the projector onto the $R_2$ representation \cite{Berman:2012vc}.
For $d=4$ to $d=6$ it is directly proportional to this projector, and we find that its trace is $ Y^{MN}{}_{MN} = 2(d-1) \mathrm{dim}\,R_2$.
For $d=7$, an additional term appears in the Y-tensor involving the antisymmetric invariant of $\Gseven$ (i.e. a projector onto also the trivial representation) and in this case $Y^{MN}{}_{MN} = 2(d-1)\mathrm{dim}\,R_2  - \mathrm{dim}\, R_1/2$. For $d=8$, the situation changes again and the trace does not have quite such a simple expression. 

The crucial information about the coset then appears in the very final term in \eqref{traceP}, which we may single out and define as
\be
r \equiv \frac{1}{2\alpha} \gM_{MN} Y^{MN}{}_{KL} \gM^{KL}\,.
\label{r}
\ee 
One finds, as summarised in table \ref{projconstants}, that for all groups except $\Geight$ the trace of the projector gives exactly the dimension of the usual $G/H$ coset minus $r$. For $\Geight$ we obtain the dimension of $\Geight/\Height$ plus $2/15$ minus $r$. 
It follows that non-zero $r$, if possible, generically corresponds to parametrisations in which there are fewer independent components of the generalised metric, signalling a coset $G/H$ of lower dimension.
Information about $H$ can be introduced in the form of a generalised vielbein, $E_M{}^{\mathcal{A}}$, with a flat index $\mathcal{A}$ transforming under $H$.
The generalised metric is then given $\gM_{MN} = E_M{}^{\mathcal{A}}E_N{}^{\mathcal{B}} \mathcal{H}_{AB}$, with the flat metric $\mathcal{H}_{AB}$ which is left invariant by local $H$ transformations.
Using the group properties of the generalised vielbein (it must preserve the Y-tensor), it is then possible to explicitly evaluate $r$, as we will see below for $\Geight$ in section \ref{8} (and for $\Gseven$ and $\Gsix$ in section \ref{usualcosets}).

\begin{table}[ht]
\centering
\begin{tabular}{lllllll}
$G$ & $\alpha$ & $\omega$ & $\gamma$ &  $\mathrm{dim} \,R_1$ & $P_{MN}{}^{MN}$ \\\hline
$\mathrm{O}(D,D)$ & 2 & 0 & 1 & $2D$ & $D^2 - r$ \\
$\Gfour$ & 3 &$ - 1/5 $ & 3 & 10 & $14 - r$ \\
$\Gfive$ & 4& $- 1/4$ & 5 & 16 & $25 - r$\\
$\Gsix$ & 6& $- 1/3$ & 10 & 27 & $42 - r$\\
$\Gseven$ & 12& $- 1/2$  & 28 & 56 & $70 - r$ \\
$\Geight$ & 60& $- 1$ & 189 & 248 & $128 + \frac{2}{15}- r$ 
\end{tabular}
\caption{Constants appearing in the projector. Here $\gamma \equiv Y^{MN}{}_{MN} / \mathrm{dim}\, R_1$ and $r$ is defined in \eqref{r}.
For the usual cosets $r = 0$ for all cases except $E_{8(8)}$, when $r = \frac{2}{15}$.
}
\label{projconstants}
\end{table}

However, this does not rule out the possibility of finding alternative parametrisations of the generalised metric which correspond to new cosets $G/H$ of lower dimension. Indeed, this underlies the non-Riemannian parametrisations of \cite{Morand:2017fnv}, which we will review from the perspective of the projector $P_{MN}{}^{KL}$ in section \ref{DFTnonRie}, and will appear below in an interesting context for the $\Geight$ ExFT.\footnote{Indeed, the general situation may be that one can extend the definition of the generalised metric such that $\mathcal{M}_{MN} =E_M{}^{\mathcal{A}}E_N{}^{\mathcal{B}} \mathcal{H}_{AB}$ where now $\mathcal{H}_{AB}$ is specified by a choice of Cartan involution of the group. We thank Martin Cederwall and the anonymous referee for making this point to us.}

\subsection{The action and equations of motion}

Let us now discuss the dynamics of the generalised metric.
Its equations of motion follow from the ExFT action, which is constructed using the requirement of invariance under the local symmetries of ExFT.
These include not only generalised diffeomorphisms but also external diffeomorphisms associated to transformations of the coordinates $X^\mu$, and various generalised gauge transformations of gauge fields that also appear in the theory. 

The projector then plays a vital role in the equations of motion for the generalised metric. (Here we are thinking only of the bosonic part of the action: if we include fermions then we will have to use a projector onto the variation of the generalised vielbein. We will comment more on this later.)
In fact, it was in this context that the projector was first written down in \cite{Berkeley:2014nza} (where it was obtained for the groups $\Gfour$ and $\Gfive$ by explicitly varying known parametrisations of the generalised metric).
When one varies the action with respect to $\gM_{MN}$, one naively obtains an expression of the form
\be
\delta S = \int \delta \gM^{MN} \mathcal{K}_{MN}\, , \qquad \mathcal{K}_{MN} \equiv \frac{\delta S}{\delta \mathcal{M}^{MN}}
\ee
but the true equations of motion are 
\be\label{eq:EoM}
P_{MN}{}^{KL} \mathcal{K}_{KL}=0\,.
\ee
The reason for this is that one must insist that the variations of the generalised metric $\delta \gM^{MN}$ are still compatible with $G$ and so we impose this by a projector. In the standard formulation of ExFT, the actions do not explicitly impose this and so one needs to include these projectors by hand though it is equivalent to just calculating the variations of the action subject to $G$-compatibility. 

Now, recalling that the projector depends on $\mathcal{M}_{MN}$, we might consider whether it is possible to find a generalised metric such that the projector vanishes:
\be
P_{MN}{}^{KL} = 0 \,,
\ee
meaning the equations of motion \eqref{eq:EoM} are trivially obeyed. This is evidently a very special possibility. It corresponds to changing the structure of the theory such that the coset is $G/G$. Furthermore, as \emph{any} variation of the generalised metric must be projected, $\delta \gM_{MN} = P_{MN}{}^{KL} \delta \gM_{KL}$, there can be no fluctuations about such a background \cite{Cho:2018alk}.\footnote{The idea of looking for generalised metrics such that $P_{MN}{}^{KL} = 0$ was originally suggested to us in this context by Diego Marqu{\'e}s.}. 

For $\mathrm{O}(d,d)$, the ``maximally non-Riemannian'' background $\cH_{MN} = \eta_{MN}$ is of this type \cite{Morand:2017fnv}. 
This background is invariant under $\mathrm{O}(d,d)$, i.e. it corresponds to a symmetric invariant tensor of the group. 
This characterisation is easy to search for in ExFT, where the symmetric product of $R_1$ with itself does not contain the trivial representation for any $\Edd$ except for $d=8$. For $\Geight$ we have $R_1 = \mathbf{248}$, which is the adjoint representation and there is an obvious symmetric quadratic invariant given by the Killing form. 
We will now discuss this ExFT and what one can say about the non-Riemannian background where the generalised metric is proportional to the Killing form.

\subsection{The $\Geight$ ExFT and its topological phase} 
\label{8}
\subsubsection*{Generalised Diffeomorphisms and the Action}

The $\Geight$ ExFT  \cite{Hohm:2014fxa} is based on an extended geometry parametrised by 248 coordinates $Y^M$ valued therefore in the adjoint of $\Geight$.
Denoting its generators as $T^M$, we define structure constants $f^{MN}{}_K$ with the convention $[T^M, T^N] = - f^{MN}{}_K T^K$, and the Killing form by
\begin{align}\label{eq:Killing}
\kappa^{MN} \equiv \frac{1}{60} \operatorname{Tr} (T^M T^N) = \frac{1}{60} f^{MP}{}_Q f^{NQ}{}_P\,.
\end{align}
We freely raise and lower all indices using $\kappa^{MN}$ and its inverse $\kappa_{MN}$.

The generalised Lie derivative of an adjoint vector of weight $\lambda$ is explicitly given by
\begin{align}
\mathcal{L}_\Lambda V^M = \Lambda^K \partial_K V^M - 60 {\left( \mathbb{P}_{\mathbf{248}} \right)}^M{}_K{}^N{}_L \partial_N \Lambda^LV^K + \lambda(V) \partial_N \Lambda^N V^M
\label{E8gld}
\end{align}
in which we have used the projector onto the adjoint representation ${\left( \mathbb{P}_{\mathbf{248}} \right)}^M{}_K{}^N{}_L$ defined by
\begin{align}
{\left( \mathbb{P}_{\mathbf{248}} \right)}^M{}_K{}^N{}_L = \frac{1}{60} f^M{}_{KP} f^{PN}{}_L.
\end{align}
Alternatively, one can write the part of this transformation involving $\Lambda^M$ in the form \eqref{gldY} involving the Y-tensor, given here by
\be
Y^{MN}{}_{KL} = - f^M{}_{LP} f^{PN}{}_K + 2 \delta^{(M}_K \delta^{N)}_L \,.
\ee
A special feature of the $\Geight$ ExFT is that it includes additional gauge transformations which appear alongside the conventional generalised Lie derivative. 
Under this extra gauge symmetry, generalised vectors transform as
\be
\delta_\Sigma V^M = - \Sigma_L f^{LM}{}_N V^N \,,
\label{constrainedgaugetransformation}
\ee
where the gauge parameter $\Sigma_M$ is not an arbitrary covector but is constrained as part of the section condition of the $\Geight$ ExFT.
This section condition applies to any two quantities $F_M$, $F^\prime_M$ which are said to be ``covariantly constrained'' meaning that they vanish when their tensor product is projected into the $\mathbf{1 \oplus 248 \oplus3875} \subset \mathbf{248 \otimes 248}$, i.e.
\be
\kappa^{MN} F_M \otimes F^\prime_N  = 0\,,\quad
f^{MNK} F_N \otimes F^\prime_K  = 0\,,\quad
{(\mathbb{P}_{\mathbf{3875}})}^{KL}{}_{MN} F_K \otimes F^\prime_L = 0\,.
\label{E8section}
\ee
These quantities include derivatives, $\partial_M$, as usual, the gauge parameters $\Sigma_M$, and a number of other gauge parameters and field \cite{Hohm:2014fxa}. 

This section condition guarantees closure of the algebra of the combined action of generalised diffeomorphisms and constrained $\Sigma_M$ transformations, which we denote by
\be
\mathbb{L}_{( \Lambda, \Sigma )} \equiv \mathcal{L}_\Lambda + \delta_\Sigma\,.
\ee 
The inclusion of the $\Sigma_M$ transformations is in fact necessary for closure: the algebra based on the ordinary generalised Lie derivative \eqref{E8gld} alone cannot be made to close on its own.
The underlying physical reason for the extra gauge transformation \eqref{constrainedgaugetransformation} is the appearance of dual graviton degrees of freedom in the generalised metric of the $\Geight$ ExFT.
For further details on these subtleties, we refer the reader to the original paper \cite{Hohm:2014fxa} or the recent review \cite{Hohm:2018qhd}.

We proceed to discuss the field content of the theory. This consists of the generalised metric, $\gM_{MN}$, an external metric, $g_{\mu\nu}$, and a pair of gauge fields $(\Aa_\mu{}^M , \Ab_{\mu M})$, with $\Ab_{\mu M}$covariantly constrained as in \eqref{E8section}.
These gauge fields have field strengths $(\mathcal{F}_{\mu\nu}{}^M,\mathcal{G}_{\mu\nu M})$ whose precise forms can be found in \cite{Hohm:2014fxa}. All these fields depend on the three-dimensional coordinates $X^\mu$ as well as the 248-dimensional coordinates $Y^M$, subject to the section condition. The gauge field $\Aa_\mu{}^M$ can be thought of as serving as a gauge field for generalised diffeomorphisms while $\Ab_{\mu M}$ is a gauge field for the constrained $\Sigma_M$ transformations. We define an improved derivative $D_\mu \equiv \partial_\mu - \mathbb{L}_{( \Aa_\mu, \Ab_\mu)}$ which is used in place of $\partial_\mu$.
The action for the $\Geight$ ExFT is constructed in \cite{Hohm:2014fxa} and is given by
\be
\begin{split}
S & = \int \textrm{d}^{3}x \textrm{d}^{248} Y   \sqrt{|g|} \,\Bigg(
\hat R[g] +  \frac{1}{240}  g^{\mu \nu} D_\mu \mathcal{M}_{MN} D_\nu \mathcal{M}^{MN}
- V( \mathcal{M},g ) + \frac{1}{\sqrt{|g|}} \mathcal{L}_{CS}
\Bigg) 
\end{split}
\label{actionE8}
\ee
where $\hat R[g]$ is the usual Ricci scalar for the metric $g_{\mu\nu}$, except constructed in terms of $D_\mu$ instead of $\partial_\mu$. 
The two terms at the end are:
\begin{align}
\begin{aligned}
	V (\mathcal{M}, g) & = - \frac{1}{240} \mathcal{M}^{MN} \partial_M \mathcal{M}^{KL} \partial_N \mathcal{M}_{KL} + \frac{1}{2} \mathcal{M}^{MN} \partial_M \mathcal{M}^{KL} \partial_L \mathcal{M}_{NK}\\
	& \qquad + \frac{1}{7200} f^{NQ}{}_P f^{MS}{}_R \mathcal{M}^{PK} \partial_M \mathcal{M}_{QK} \mathcal{M}^{RL} \partial_N \mathcal{M}_{SL}\\
	& \qquad - \frac{1}{2} \partial_M \ln |g| \partial_N \mathcal{M}^{MN} - \frac{1}{4} \mathcal{M}^{MN} \left(  \partial_M \ln |g| \partial_N \ln |g| + \partial_M g^{\mu \nu} \partial_N g_{\mu \nu} \right)\,,
\label{V8}
\end{aligned}
\end{align}
which is usually referred to as the ``potential'', taking the point of view of the external three-dimensional space,
and the Chern-Simons term:
\begin{align}
S_{\text{CS}} \sim \int_{\Sigma^4} \textrm{d}^4x \int \textrm{d}^{248} Y \left( \mathcal{F}^M \wedge \mathcal{G}_M - \frac{1}{2} f_{MN}{}^K \mathcal{F}^M \wedge \partial_K \mathcal{G}^{\mathcal{N}} \right)
\end{align}
written here in a manifestly gauge invariant form using the usual construction of an auxiliary space $\Sigma^4$  whose boundary $\partial \Sigma^4$ is the physical three-dimensional space, and where $\wedge$ denotes the usual product with respect to the external indices, $\mu,\nu,\dots$.

\subsubsection*{Generalised metric and projector}

Conventionally, we view the generalised metric as being an element of $\Geight / H$, with $H=\mathrm{SO}(16)$, and then this coset is parametrised in terms of a spacetime metric and $p$-form fields. Instead, following the intuition from the DFT approach of \cite{Morand:2017fnv} where the generalised metric was defined as a symmetric two index object obeying the $\mathrm{O}(d,d)$ compatibility condition \eqref{comptcon}, we will define the $\Geight$ generalised metric by the properties that are needed in \cite{Hohm:2014fxa} to ensure the invariance of the action \eqref{actionE8}.
Thus we define the $\Geight$ generalised metric to be the symmetric two index object that obeys the constraints:
\be
\gM_{MK} \gM_{NL} \gM_{PQ} f^{KLQ} = - f_{MNP} \,,\quad \gM_{MK} \kappa^{KL} \gM_{LN} = \kappa_{MN} \, .
\label{8Mdef}
\ee
One can check that the conventional coset parametrisation of $\gM_{MK}$ obeys these constraints but new results will follow from a solution to these constraints that does not obey the coset parametrisation.
The full generalised Lie derivative (including the additional transformations involving $\Sigma_M$) of the generalised metric takes the form
\begin{align}
\mathbb{L}_{(\Lambda,\Sigma)} 
\mathcal{M}_{MN} = \Lambda^P \partial_P \mathcal{M}_{MN} + 2 \cdot 60 P_{MN}{}^{KL} \left(
\partial_K \Lambda^P 
 +\frac{1}{60} f^{QP}{}_{K} \Sigma_Q\right) 
\mathcal{M}_{PL}\,,
\end{align}
with the projector given simply by
\be
P_{MN}{}^{KL} =\frac{1}{60} \gM_{MQ} f^Q{}_{NP} f^{P (K }{}_R \gM^{L) R} \,.
\ee
The trace is
\be
P_{MN}{}^{MN} = \frac{1}{2} \left( \kappa^{MN} \gM_{MN} + 248 \right) \,.
\ee
Now, for the usual $\Geight / \mathrm{SO}(16)$ coset, we introduce a generalised vielbein $E_M{}^{\mathcal{A}}$ such that \cite{Baguet:2016jph}
\be
E_M{}^{\mathcal{A}} \equiv ( E_M{}^A, E_M{}^{IJ})\,,\quad \kappa^{MN} E_M{}^A E_N{}^B = \delta^{AB} \,,\quad  \kappa^{MN} E_M{}^{IJ} E_N{}^{KL} = - 2 \delta^{I[K} \delta^{L]J} \,,
\ee 
where $A$ is a spinor index corresponding to the $\mathbf{128}$ of $\mathrm{SO}(16)$, and $I$ the 16-dimensional vector representation, with $E_M{}^{IJ} = - E_M{}^{JI}$ in the $\mathbf{120}$ of $\mathrm{SO}(16)$. 
The generalised metric is then given by $\gM_{MN} = E_M{}^A E_N{}^B \delta_{AB} + \frac{1}{2} E_M{}^{IJ} E_N{}^{KL} \delta_{IK} \delta_{JL}$ and it follows from the defining properties of the vielbein that $\kappa^{MN} \gM_{MN} = 128 - 120 =8$.
Thus we find $P_{MN}{}^{MN} = 128$ as expected. 

Now we can consider whether there are alternative parametrisations of $\gM_{MN}$ such that $P_{MN}{}^{MN} \neq 128$. 
Remarkably, we can immediately write down a choice of $\mathcal{M}_{MN}$ such that $P_{MN}{}^{KL}$ vanishes identically, given by
\be
\gM_{MN} = - \kappa_{MN} \, .
\ee
This is easily checked to be compatible with the defining constraints \eqref{8Mdef} for $\gM_{MN}$ (no other multiple of the Killing form is).
The projector then vanishes as $f^{P (KL)} = 0$. 

\subsubsection*{Restricting to the ``topological phase''}

Now let us consider what this implies for the equations of motion.
On general grounds, as we have explained, the equations of motion of $\gM_{MN}$ itself will be of the form $P_{MN}{}^{KL} \mathcal{K}_{KL} = 0$, where $\mathcal{K}_{MN}$ is the result of varying the action with respect to $\gM^{MN}$.
As the projector vanishes for $\gM_{MN} = - \kappa_{MN}$, the equations of motion are trivially obeyed. 

Now consider the variation of the other fields in the action. 
For instance, the equation of motion of the external metric is:
\be
\begin{split}
0 & = 
\hat R_{\mu \nu} - \frac{1}{2} g_{\mu\nu} \left( \hat R[g] + \frac{1}{240}  g^{\rho\sigma} D_\rho \mathcal{M}_{MN} D_\sigma \mathcal{M}^{MN} - V( \mathcal{M},g )\right) 
\\ & \qquad
+ \frac{1}{240} D_\mu \gM_{MN} D_\nu \gM^{MN} 
+ \frac{1}{2}\sqrt{|g|}^{-1} g_{\mu\nu} \partial_M \left( \sqrt{|g|} ( \partial_N \gM^{MN} + \gM^{MN} \partial_N \ln |g| )\right) 
\\ & \qquad
- \frac{1}{2} \sqrt{|g|}^{-1} \partial_M ( \sqrt{|g|} \gM^{MN} ) \partial_N g_{\mu\nu}  
- \frac{1}{2} \gM^{MN} g_{\mu \rho} \partial_M g^{\rho \sigma} \partial_N g_{\sigma \nu} 
- \frac{1}{2} \gM^{MN} \partial_M \partial_N g_{\mu\nu}
 \,.
\end{split} 
\ee 
Here $\hat R_{\mu\nu}$ is defined to be the result of varying $\hat R[g]$ with respect to $g_{\mu\nu}$. 
Now, when $\gM_{MN} = - \kappa_{MN}$ all terms involving the generalised metric vanish identically, either because $D_\mu \kappa_{MN} = 0$ (as the generalised Lie derivative appearing in the definition of $D_\mu$ preserves the Killing form) or because of the section condition $\kappa^{MN} \partial_M \otimes \partial_N  = 0$. 
Similarly, the equations of motion of the gauge fields $\Aa_{\mu}{}^M$, $\Ab_{\mu M}$ will involve $\gM_{MN}$ only in the form of (derivatives of) $D_{\mu} \gM_{MN}$, and so the contribution of the generalised metric to these equations of motion also vanishes identically. 

We can conclude that the equations of motion for $(g_{\mu\nu}, \Aa_{\mu}{}^M, \Ab_{\mu M})$ when $\gM_{MN} = - \kappa_{MN}$ are those that are obtained from the truncation of the ExFT action obtained by setting $\gM_{MN}
 = - \kappa_{MN}$ within the action, i.e. in this background the dynamics of the resulting fields are governed by:
\begin{align}
\begin{aligned}
S  = \int \textrm{d}^3x \,\textrm{d}^{248} Y \sqrt{|g|}  \hat R[g] + \int_{\Sigma^4} \textrm{d}^4x \,\textrm{d}^{248} Y \left( \mathcal{F}^M \wedge \mathcal{G}_M - \frac{1}{2} f_{MN}{}^K \mathcal{F}^M \wedge \partial_K \mathcal{G}^{\mathcal{N}} \right)\,.
\end{aligned}
\label{truncaction}
\end{align}
Now, ordinary three-dimensional gravity is topological, so this action naively resembles that of a topological gravity theory plus a Chern-Simons term, though matters are complicated by the dependence on the coordinates $Y^M$ and the modified partial derivative used in the construction of the Ricci scalar.
Remarkably, however, the entire truncation \eqref{truncaction} including the external metric and the gauge fields is indeed a novel sort of topological theory.
This was shown in \cite{Hohm:2018ybo} where the theory described by the action \eqref{truncaction} was reformulated as a Chern-Simons theory based on a ``Leibniz algebra''\footnote{A generalisation of a Lie algebra in which the product (replacing the Lie bracket) is not necessarily antisymmetric.} incorporating both the three-dimensional Poincar\'e symmetry and the generalised diffeomorphisms of the $\Geight$ ExFT. (To think of this theory as being topological in the three-dimensional sense, we can view the gauge group of the Chern-Simons theory as being infinite dimensional due to the dependence on the $Y^M$ coordinates, while the integration over these coordinates in the action is part of the definition of an inner product on this infinite dimensional gauge group).
This was termed the ``topological phase'' of the $\Geight$ ExFT, and was achieved by the perhaps ad hoc elimination of the generalised metric by setting $\gM_{MN} = 0$. This was motivated by a desire to eliminate the degrees of freedom in the generalised metric while maintaining unbroken $\Geight$ (a truncation to the more natural vacuum $\gM_{MN} = \delta_{MN}$ would break $\Geight$ to $\mathrm{SO}(16)$, for example). 

We propose that the true, non-singular origin of the topological phase is in fact the maximally non-Riemannian background $\gM_{MN} = -\kappa_{MN}$. 
We expect that this can be consistently viewed as taking the defining coset to be $\Geight/\Geight$, with no internal bosonic degrees of freedom. 
It is interesting to realise that the consistency of this truncation depends crucially on the fact that $\kappa^{MN} \partial_M \otimes \partial_N = 0$ by the section conditions of the $\Geight$ ExFT. Thus, the remaining fields in \eqref{truncaction} may still depend on the extended coordinates $Y^M$ subject to this constraint.

Let us make a short comment about the fermions of the $\Geight$ ExFT. 
We would expect that after truncating the generalised metric degrees of freedom that we should also truncate out the internal fermions. 
At this point the supersymmetry of the non-Riemannian background is a little mysterious since usually in ExFT the fermions should transform in a representation of $H$. What this means when $H=\Geight$ is uncertain but what is apparent is that one cannot just naively insert the condition $\gM_{MN} = -\kappa_{MN}$ into the generalised Killing spinor equations. The realisation of fermions in the non-Riemannian background has yet to be determined. Note that the variation of the action with respect to the generalised vielbein, $E_M{}^A$, requires a projector to ensure that $\delta E_M{}^A$ is not arbitrary. Evidently this projector will depend explicitly on the precise form of $H$ (whereas the projector $P_{MN}{}^{KL}$ acting on variations of the generalised metric only knew about $H$ implicitly, through the term $\gM_{MN} Y^{MN}{}_{KL} \gM^{KL}$) and so must be constructed on a case-by-case basis when starting from a particular non-Riemannian parametrisation of $\gM_{MN}$.

A related technical comment is to note that setting $\gM_{MN} = - \kappa_{MN}$ is consistent with the invariance of the ExFT action under external diffeomorphisms with parameter $\xi^\mu(X,Y)$, which includes a generalised metric dependent transformation of $\Aa_\mu{}^M$, namely
\be
\delta_\xi \Aa_\mu{}^M \supset \gM^{MN} g_{\mu\nu} \partial_N \xi^\nu \,.
\label{thistransf}
\ee
Normally, this requires cross-cancellation between the scalar potential and the other parts of the action. If this vanishes, $V(\gM = -\kappa, g) = 0$, then one might be concerned whether the action is still invariant. 
However, when one inspects the calculation in \cite{Hohm:2014fxa} of the variation of the action under these transformations, one finds that all possible terms that could spoil invariance vanish by the section condition on setting $\gM^{MN} = - \kappa^{MN}$.

\section{Non-Riemannian backgrounds in $\mathrm{O}(D,D)$ DFT}
\label{DFTnonRie}

In this section we first revisit the possible parametrisations of $\mathrm{O}(D,D)$ generalised metrics from the perspective of the coset projector. 
We demonstrate how the classification of $\mathrm{O}(D,D)$ non-Riemannian parametrisations of Morand and Park \cite{Morand:2017fnv} fits into this picture.
Then, we will review the explicit details of these parametrisations and look at some examples which will inspire us in our later study of the $\Gfour$ ExFT.

\subsection{Generalised metric and coset projectors}

Let us first recall that the generalised metric of DFT may be defined as a symmetric matrix $\cH_{MN}$ obeying the compatibility condition $\cH_{MK} \eta^{KL} \cH_{LN} = \eta_{MN}$ with the $\mathrm{O}(D,D)$ structure.
It transforms under $\mathrm{O}(D,D)$ generalised diffeomorphisms generated by a generalised vector $\Lambda^M = ( \Lambda^i , \lambda_i )$ according to the generalised Lie derivative \eqref{gldY} with the Y-tensor $Y^{MN}{}_{PQ} = \eta^{MN} \eta_{PQ}$ and $\omega = 0$.
The $\mathrm{O}(D,D)$ section condition $\eta^{MN} \partial_M \otimes \partial_N = 0$ may be solved by $\partial_i \neq 0 , \tilde \partial^i = 0$, where the doubled coordinates are $Y^M = (Y^i, \tilde Y_i)$,
After solving the section condition in this way, generalised diffeomorphisms produce $D$-dimensional diffeomorphisms generated by $\Lambda^i$ and $B$-field gauge transformations with parameter $\lambda_i$.
This leads to the usual parametrisation given in \eqref{Hnormal} in terms of the spacetime metric, $g_{ij}$, in string frame, and the $B$-field. The generalised dilaton may then be identified as $e^{-2\gendil} = e^{-2\Phi} \sqrt{|g|}$, where $\Phi$ is the spacetime dilaton. There is an implicit assumption in \eqref{Hnormal} that the $D \times D$ block $\cH^{ij}$, which is identified with the inverse spacetime metric, is invertible. 

The $\mathrm{O}(D,D)$ compatibility condition implies the existence of two projectors
\be
P_M^N = \frac{1}{2} ( \delta_M^N + \eta^{NP} \cH_{PM} )\,,\quad \bar P_M^N = \frac{1}{2} ( \delta_M^N - \eta^{NP} \cH_{PM} )\, ,
\ee
such the projector $P_{MN}{}^{KL}$, that appears in the generalised Lie derivative of the generalised metric \eqref{lgm}, factorises as 
\be
P_{MN}{}^{KL} = 2 P_M^{(K} \bar P_N{}^{L)} \,.
\ee 
In the usual parametrisation \eqref{Hnormal}, the trace $\eta^{MN} \cH_{MN}$ is zero, and hence $P_{MN}{}^{MN} = D^2$, as expected for the $\mathrm{O}(D,D) / \mathrm{O}(D) \times \mathrm{O}(D)$ coset.

Let us suppose instead that the trace is not necessarily zero. 
Then, as $P_M^N$ and $\bar P_M^N$ are still projectors, we can have $\eta^{MN} \cH_{MN} = 2 y$, for some integer $y$, with $-D\leq y \leq D$,
such that $P_M^M = D + y$, $\bar P_M^M = D - y$.

We can define ``square roots'' of the projectors, namely matrices $\VR_{M A}$ and $\VL_{M \bA}$, where $A = 1 , \dots , D+y$, $\bA = 1 , \dots D-y$.
These obey 
\be
\VR_{M A } h^{A B  } \VR_{N B  } = \frac{1}{2} ( \cH_{MN} + \eta_{MN} )\,,\quad
\VR_{MA } \eta^{MN} \VR_{N B  } = h_{A B  } \,,\quad
\cH^{MN} \VR_{N A } = \eta^{MN} \VR_{NA } \,,
\label{viel1}
\ee
\be
\VL_{M{\bA}} \VL_{N{\bB}} \bh^{\bA \bB} = \frac{1}{2} ( \cH_{MN} - \eta_{MN} )\,,\quad
\VL_{M\bA } \eta^{MN} \VL_{N \bB  } = -\bh^{\bA \bB  } \,,\quad
\cH^{MN} \VL_{N \bA} = - \eta^{MN} \VL_{N \bA} \,,
\label{viel2}
\ee
where $h_{A B}$ and $\bh_{\bA \bB}$ are respectively $(D+y) \times (D+y)$ and $(D-y) \times (D-y)$ diagonal matrices of signatures $(p,q)$ and $(\bar p, \bar q)$. This is quite general; we will see how different choices of signature allow for different coset descriptions and constrains $(p,q)$ and $(\bar p, \bar q)$.
Constructing a vielbein for the full generalised metric,
\be
E_M{}^{\mathcal{A}} = ( \VR_M{}^A , \VL_M{}^{\bar A} ) \,,\quad \cH_{MN} = E_M{}^{\mathcal{A}} E_N{}^{\mathcal{B}} \cH_{\cal{A}\cal{B}} \,,
\ee
where the $2D \times 2D$ flat metric,
\be
\cH_{\cal{A} \cal{B}} \equiv \begin{pmatrix}
h_{A B} & 0 \\ 
0 & \bar h_{\bA \bB}
\end{pmatrix} \,,
\ee
is of signature $(p+\bar p, q + \bar q)$ we can check that
\be
\eta^{\mathcal{A} \mathcal{B} } \equiv E_M{}^{\mathcal{A}} E_N{}^{\mathcal{B}} \eta^{MN} = \begin{pmatrix}
h^{A B} & 0 \\ 
0 &- \bar h^{\bA \bB}
\end{pmatrix} \,
\ee
then has signature $(p+\bar q,q+\bar p)$.
Now, $E_M{}^{\mathcal{A}}$ must be an $\mathrm{O}(D,D)$ group element. This means that $\eta^{\cal{A}\cal{B}}$ should have signature $(D,D)$ and so be equivalent (by a choice of basis for the flat indices) to $\eta^{MN}$.
Hence the only possibilities obey $p+\bar q = D$, $q+\bar p = D$. This means that $p-\bar p = q - \bar q = y$ which is consistent with the trace being $\eta^{MN} \cH_{MN} = \eta^{\cal{A} \cal{B}} \cH_{\cal{A}\cal{B}} = p+q - \bar p - \bar q = 2y$. 

The conclusion is that when $\eta^{MN} \cH_{MN} = 2y$, the allowed denominator groups are $H = \mathrm{O}(p,q) \times \mathrm{O}(p-y,q-y)$ with $p+q-y=D$ and $-D \leq y \leq D$.
To connect with the results of \cite{Morand:2017fnv}, we can trade the integer $y$ for a pair of non-negative integers $(n, \bar n)$ such that $y = n - \bar n$. 
We also let $t = p - n$, $s=q-n$, and $d=t+s$, such $D= d+ n+\bar{n}$.
Then instead of the usual $\mathrm{O}(D,D)/\mathrm{O}(D) \times \mathrm{O}(D)$ coset we have
\be
\frac{ \mathrm{O}(d+n+\bar{n}, d+n+\bar{n} ) }{\mathrm{O}(t+n,s+n) \times \mathrm{O}(t+\bar{n} , s+\bar{n} )} \,.
\label{ODDcoset}
\ee
The denominator agrees with the generalised Lorentz factors established in \cite{Morand:2017fnv}.
Note that this coset has dimension $d^2 + 2d(n+\bar n) + 4 n \bar n = D^2 - (n-\bar{n})^2$. There are thus $(n-\bar{n})^2$ fewer components than would ordinarily be present.


Note that the explicit parametrisation that will be used in the subsequent subsection does not make this component counting manifest, as it uses variables which are written in a $D$-dimensionally covariant manner. As a result, there are shift symmetries present (see \eqref{dftshift} below) which complicate the choice of what should be regarded as the true independent variables. This suggests there ought to be an alternative formulation which exhibits the coset structure \eqref{ODDcoset} more clearly.\footnote{We thank Jeong-Hyuck Park for detailed discussions on this issue and for sharing an alternative derivation of the fact there are $D^2-(n-\bar n)^2$ independent components.}

\subsection{Review of Morand-Park classification}
\label{MorandPark}

Dropping the assumption of the invertibility of the $D \times D$ block $\cH^{ij}$ in the normal parametrisation \eqref{Hnormal} led to the classification of 
$\mathrm{O}(D,D)$ generalised metrics in \cite{Morand:2017fnv}. Taking the section condition solution, $\partial_i \neq 0$, $\tilde \partial^i = 0$, they found that the most general parametrisation of the generalised metric is given by
\be
\cH_{MN} = \begin{pmatrix} 1 & B \\ 0 & 1 \end{pmatrix} 
\begin{pmatrix}
K_{ij} & X_i^a Y^j_a - \bX_i^{\ba} \bY^j_{\ba} \\ 
X_j^a Y^i_a - \bX_j^{\ba} \bY^i_{\ba} & H^{ij} 
\end{pmatrix} 
\begin{pmatrix}
1 & 0 \\ -B & 1
\end{pmatrix} \,.
\label{HnonRie}
\ee
Here both $H^{ij}$ and $K_{ij}$ are symmetric $D \times D$ matrices which may be non-invertible, with $\{X, \bX \}$ spanning the kernel of $H^{ij}$ and $\{ Y , \bY \}$ spanning the kernel of $K_{ij}$.
Both kernels have dimensions $n+\bar{n}$, and we index the zero vectors by $a=1,\dots,n$ and $\ba =1 ,\dots \bn$.
Explicitly, 
\be
H^{ij} X_j^a = 0 \,,\quad
H^{ij} \bX_j^{\ba} = 0 \,,\quad
K_{ij} Y^j_a = 0 \,,\quad
K_{ij} \bY^j_{\ba} = 0 \,.
\ee
We have some completeness relations which are necessary for the invertibility of $\cH_{MN}$, namely
\be
H^{ik} K_{kj} + Y^i_a X_j^a + \bY^i_{\ba} \bX_j^{\ba} = \delta^i_j\,,
\quad
Y^i_a X_i^b = \delta_a^b \,,\quad \bY^i_{\ba} \bX_i^{\bb} = \delta_{\ba}^{\bb} 
\,,\quad
Y_a^i \bX_i^{\bb} = 0 = \bY^i_{\ba}X_i^b  \,,
\ee
which imply $H^{ik} K_{kl} H^{lj} = H^{ij}$, $K_{ik} H^{kl} K_{lj} = K_{ij}$. 
These objects are all tensors under diffeomorphisms and invariant under $B$-field gauge transformations.
We see that the trace of the generalised metric is no longer zero, but given by $\cH^M{}_M = 2(n-\bn)$, in agreement with the analysis of the previous subsection, with $0 \leq n+\bn \leq D$.
Note that ${X,\bX}$ and ${Y,\bY}$ are a preferred basis for the zero vectors of $H$ and $K$. 
Any other basis $X^{\prime u}_i$, $Y^{\prime i}_{u}$, where $u=1,\dots n+\bar{n}$, would be such that
\be
Z_i{}^j \equiv  X_i^a Y^j_a - \bX_i^{\ba} \bY^j_{\ba} = X^{\prime u}_i \sigma_u{}^v Y^{\prime j}{}_v
\ee
where $\sigma_u{}^v$ is conjugate to $\mathrm{diag}( \delta^a_b , - \delta^{\bar{a}}_{\bar{b}} )$. Thus ${X,\bX}$ and ${Y,\bY}$ diagonalise $\sigma_u{}^v$.
Finally, note there is also a shift symmetry preserving the parametrisation \eqref{HnonRie}, involving arbitrary parameters $b_{i a}$, $\bar b_{i \bar a}$:
\be
\begin{split}
Y_{a}^{i }  \rightarrow&   Y_{a}^{i }+H^{i j }b_{j  a}\,,\\
\bY_{\bar{a} }^{i }  \rightarrow&  \bar Y_{\bar{a} }^{i }+H^{i j }\bar{b}_{j \bar{a} }\,,\\
K_{i j }  \rightarrow&   K_{i j }
	-2X^{a}_{(i }K_{j )k}H^{kl}b_{l a}
	-2\bX^{\bar{a} }_{(i }K_{j )k}H^{kl}\bar{b}_{l\bar{a} }
	+(X_{i }^{a}b_{k a}	+\bX_{i }^{\bar{a} }\bar{b}_{k\bar{a} })H^{kl}(X_{j }^{b}b_{l b}+\bX_{j }^{\bar{b}}\bar{b}_{l\bar{b}})\,,\\
B_{i j }  \rightarrow&   B_{i j }
	-2X^{a}_{[i }b_{j ]a}+2\bX^{\bar{a} }_{[i }\bar{b}_{j ]\bar{a} }
	+2X^{a}_{[i }\bX^{\bar{a} }_{j ]}\left(Y_{a}^{k}\bar{b}_{k\bar{a} }
	+\bar Y_{\bar{a} }^{k}b_{k a}+b_{k a}H^{kl}\bar{b}_{l\bar{a} }\right)\,,
\label{dftshift}
\end{split}
\ee
which we can view as eliminating some components of the $B$-field in the non-Riemannian geometry.

A variety of interesting example have been considered in \cite{Morand:2017fnv}. For instance, 
$(n,\bn) = (D,0)$ corresponds to the maximally non-Riemannian case, $\cH_{MN} = \eta_{MN}$. 
When $n = \bar{n}$ the parametrisations may be connected by $\mathrm{O}(D,D)$ transformations to Riemannian parametrisations. An example, which we will discuss below, is the $(1,1)$ non-Riemannian metric corresponding to the Gomis-Ooguri limit of string theory, or to the T-dual of a supergravity solution. 
The case $(n,\bn) = (D-1,0)$ gives an ultra-relativistic (Carroll) geometry, while $(n,\bn) = (1,0)$ or $(0,1)$ provides a version of non-relativistic Newton-Cartan geometry. (In this case, the transformation \eqref{dftshift} in fact reduces to known non-relativistic transformations termed Milne transformations or Galilean boosts \cite{Morand:2017fnv}.)
In general, the non-Riemannian background  \eqref{HnonRie} can be studied using the doubled sigma model, and it was shown in \cite{Morand:2017fnv} that the zero vectors $X_i{}^a$ pick out $n$ string target space coordinates which become chiral, while the $\bar X_i{}^{\bar a}$ lead to $\bar n$ antichiral directions.

The paper \cite{Morand:2017fnv} also introduced generalised vielbeins as follows. 
Let $d$ denote the rank of $H^{ij}$ and $K_{ij}$, such that $D=d+n+\bar{n}$. 
Suppose that $H^{ij}$ and $K_{ij}$ have signature $(t,s, n+\bar n)$, and define the flat matrices
\be
h_{A B  } = \begin{pmatrix} \eta_{mn} & 0 & 0 \\ 0 & - \delta_{ab} & 0 \\ 0 & 0 & \delta_{ab} \end{pmatrix} \,,\quad
\bh_{\bA \bB} = \begin{pmatrix} \eta_{\bar m \bar n} & 0 & 0 \\ 0 & - \delta_{\ba\bb} & 0 \\ 0 & 0 & \delta_{\ba\bb} \end{pmatrix}\,,
\ee
for the $\mathrm{O}(t+n,s+n)$ and $\mathrm{O}(t+\bar n, s+\bar n)$ factors respectively. 
We have two types of flat indices, one for each factor, which we write as $A  = ( m, a, a)$ where $m=1,\dots,D-n-\bar{n}$ and $a=1,\dots,n$, and $\bA  = ( \bar m, \bar a, \bar a)$ where $m=1,\dots,D-n-\bar{n}$ and $\bar a=1,\dots,\bar n$. The matrices $\eta_{mn}$ and $\eta_{\bar m \bar n}$ are separate copies of the Minkowski metric of signature $(t,s)$. 
Using these, we can introduce vielbeins for the degenerate matrices $K$ and $H$:
\be
K_{ij} = k_i{}^{m} k_j{}^{n} \eta_{mn} = \bk_i{}^{\bm} \bk_j{}^{\bn} \eta_{\bm\bn} \,,\quad
H^{ij} = h^i{}_{m} h^j{}_{n}\eta^{mn} = \bh^i{}_{\bm} \bh^j{}_{\bn}\eta^{\bm\bn} \,,
\ee
which obey
\be
X_i^a h^i{}_{m} = \bX_i^a h^i{}_m = 0 = Y^i_a k_i{}^{m}  = \bY^i_a k_i{}^m \,,\quad
h^{i}{}_{m} k_{i}{}^{n} = \delta_{m}^{n}\,,\quad
h^{i}{}_{m} k_{j}{}^{m} + X_j^a Y^i_a + \bX_j^{\ba} \bY^i_{\ba} = \delta^i_j
\ee
and similarly for the barred quantities.
Now define
\be
k_i{}^A  = \begin{pmatrix}  k_i{}^{m} & X_i^{a} & X_i^{a} \end{pmatrix} \,,\quad
h^i{}_A  = \begin{pmatrix}  h^i{}_{m} & Y^i_{a} & Y^i_{a} \end{pmatrix} \,,
\ee
\be
\bar k_i{}^{\bA} = \begin{pmatrix}  \bar k_i{}^{\bmu} & \bar X_i^{\bar a} & \bar X_i^{\bar a} \end{pmatrix} \,,\quad
\bar h^i{}_{\bA} = \begin{pmatrix}  \bar h^i{}_{\bmu} & \bar Y^i_{\bar a} & \bar Y^i_{\bar a} \end{pmatrix} \,,
\ee
out of which we construct 
\be
\VR_{M A } = \frac{1}{\sqrt{2}} \begin{pmatrix} k_i{}_{A} + B  _{ij} h^{j}{}_A  \\ h^i{}_A  \end{pmatrix} \,,\quad
\VL_{M \bA} = \frac{1}{\sqrt{2}} \begin{pmatrix} 
	- \bar k_{i \bA} + B_{ij} \bar h^{j}{}_{\bA} \\
	\bar h^{i}{}_{\bA} 
\end{pmatrix} \,,
\ee
obeying \eqref{viel1} and \eqref{viel2} as required.

\subsection{Examples: Gomis-Ooguri limit and timelike duality}
\label{DFTGO}

Here we review two closely linked examples of DFT non-Riemannian geometry.

\subsubsection*{Gomis-Ooguri}

The original idea of Gomis-Ooguri \cite{Gomis:2000bd} is to consider the string sigma model in a special background for which one take a certain scaling limit leading to a description of string theory in a non-relativistic background geometry. This limit can be taken starting with the flat background
\be
ds^2 = G(-dt^2 + dz^2) + d\vec{x}_8^2 \,,\quad B = (G-\mu) dt \wedge dz \,,
\label{GOstart}
\ee
where $G$ and $\mu$ are parameters which we can tune. The choice of the $B$-field here is vital in order to take $G \rightarrow \infty$. Although this is singular in the standard Polyakov action it is non-singular in equivalent descriptions and, in particular, in the Hamiltonian or doubled approach to the string.
This can be seen by constructing the generalised metric describing the background \eqref{GOstart} by doubling only the worldsheet directions $t$ and $z$:
\be
\cH_{MN} = \begin{pmatrix}
-2 \mu + \mu^2 G^{-1} & 0 & 0& 1 - \mu G^{-1} \\
0 & 2\mu - \mu^2 G^{-1} & 1 - \mu G^{-1}& 0 \\
0 &  1 - \mu G^{-1} & - G^{-1} & 0 \\
 1 - \mu G^{-1} & 0 & 0 & G^{-1} 
\end{pmatrix} 
\ee
which, for $G \rightarrow \infty$, is non-singular but non-Riemannian. We have 
\be
\cH_{MN} = \begin{pmatrix}
-2 \mu  & 0 & 0& 1  \\
0 & 2\mu  & 1  & 0 \\
0 &  1 & 0 & 0 \\
 1 & 0 & 0& 0
\end{pmatrix} \,,
\quad
H^{ij} = 0 \,,\quad K_{ij} = 0 \,,\quad B_{ij} = - \mu \begin{pmatrix} 0 & 1 \\ -1 & 0 \end{pmatrix} \,,
\label{GOH}
\ee
and the preferred basis of zero vectors is
\be
X_i = \frac{1}{\sqrt{2}} \begin{pmatrix} 1\\ 1 \end{pmatrix} \,,\quad
Y^i = \frac{1}{\sqrt{2}} \begin{pmatrix} 1 \\ 1 \end{pmatrix} \,,\quad
\bX_i = \frac{1}{\sqrt{2}} \begin{pmatrix} 1 \\ -1 \end{pmatrix} \,,\quad
\bY^i = \frac{1}{\sqrt{2}} \begin{pmatrix} 1 \\ -1 \end{pmatrix} \,.
\ee

\subsubsection*{Non-Riemannian geometry from timelike duality}
\label{DFTtimelike}

We can also obtain a non-Riemannian generalised metric by acting with T-duality on the supergravity solution corresponding to a fundamental string solution.
This appeared in the DFT context first in \cite{Lee:2013hma}, although the timelike dual of the F1 solution was studied long ago in \cite{Welch:1994qm}. The F1 solution is: 
\be
\begin{split}
ds^2  = H^{-1} ( - dt^2 + dz^2 ) + d\vec{x}_8{}^2 \,,\quad
B_{tz}  = H^{-1} + c \,,\quad
e^{-2\phi} & = H \,,\quad
H  = 1 + \frac{h}{r^6} \,,\quad r \equiv |\vec{x}_8|
\end{split} 
\ee
Normally one takes $c = -1$ such that the B-field vanishes at infinity: in general it should lie in the range $0 > c > -2$ \cite{Park:2015bza}. 
After constructing the doubled generalised metric and dilaton using the usual parametrisation \eqref{Hnormal}, we can T-dualise in both the $t,z$ directions, giving the T-dual generalised metric (the part of the generalised metric describing the transverse space with coordinates $\vec{x}_8$ is trivial so we do not write it)
\be
\tilde{\cH}_{MN}  = \begin{pmatrix} 
-H& 0 & 0 & 1+cH \\ 0 & H & 1+cH &0 \\ 0 & 1+cH & 2c + c^2H  & 0 \\ 1+cH & 0 & 0 & -2c - c^2H \end{pmatrix} \,,
\label{negF1H}
\ee
while the generalised dilaton is invariant and is $e^{-2 \gendil} = 1$.
Defining
\be
\tilde H \equiv - c(2+cH) = - ( c^2+2c) - \frac{c^2 h}{r^6}\,,
\ee
the corresponding spacetime geometry is
\be
\begin{split}
ds^2  =\tilde H^{-1}( - d \tilde t^2 + d \tilde z^2 ) + d\vec{x}_8{}^2 \,,\quad
B_{\tilde t \tilde z} = -\tilde H^{-1} + c^{-1} \,,\quad
e^{-2\phi}  = |\tilde H| \,.
\end{split} 
\ee
The ADM mass of the solution is $M_{ADM} = \frac{R_{\tilde z}}{l_s^2} \frac{c}{c+2}$ \cite{Park:2015bza}.
Let us focus on what happens for the special values $c=-1$ and $c=0$.
\begin{itemize}
\item $c=-1$ corresponds to the usual asymptotically flat F1 solution, and gives rise here to a dual solution which is automatically asymptotically flat with $\tilde H = 2 -H$ being given by
\be
\tilde H = 1 - \frac{h}{r^6} \,.
\ee
The ADM mass is \emph{minus} that of the F1, and the solution can be interpreted as in \cite{Dijkgraaf:2016lym} as describing a \emph{negative tension F1}, or negative F1 for short. There is a singularity at $\tilde H=0$, which can be thought of as marking the position of a ``bubble'' surrounding the negative tension brane, inside which the spacetime signature flips and we should use an exotic variant of string theory, of the type investigated by Hull \cite{Hull:1998vg,Hull:1998ym}, to describe its physics.
As the generalised metric is non-singular at $\tilde H = 0$, the DFT description is perfectly well-defined (see \cite{Hohm:2011dv, Blair:2016xnn} for some discussion of such exotic theories in DFT). Indeed, we see that in this case the generalised metric is
\be
\tilde \cH_{MN} = \begin{pmatrix} \tilde H - 2 & 0 & 0 & \tilde H -1 \\ 0 & 2-\tilde H & \tilde H - 1 & 0 \\ 0 & \tilde H - 1 & \tilde H & 0 \\ \tilde H - 1 & 0 & 0 & \tilde H \end{pmatrix} 
\Rightarrow
\tilde \cH_{MN}\Big|_{\tilde H = 0} = \begin{pmatrix}
-2 & 0 & 0 & -1 \\ 
0 & 2 & -1 & 0 \\
0 & -1 & 0 & 0 \\
-1 & 0 & 0 & 0 
\end{pmatrix} 
\ee
which is in fact exactly of the non-Riemannian type appearing in the Gomis-Ooguri limit, and so is described by the same parametrisation with $\mu=1$ and $(X,Y)$ interchanged with $(\bar X, \bar Y)$ (we could also change the sign of the original $B$-field). This maybe provides an interesting interpretation of the singularity in the background of a negative brane: the string theory becomes non-relativistic at the special point $\tilde H = 0$.

\item $c=0$ corresponds to $\tilde H=0$ and the spacetime is singular. We see though that the generalised metric \eqref{negF1H} is well-defined, and given by
\be
\tilde{\cH}_{MN} = \begin{pmatrix} 
-H& 0 & 0 & 1 \\ 0 & H & 1 &0 \\ 0 & 1 & 0  & 0 \\ 1 & 0 & 0 & 0 \end{pmatrix} \,.
\label{Hnrex}
\ee
which clearly describes the same type of non-Riemannian background as the Gomis-Ooguri limit, identifying $H = + 2\mu$.

\end{itemize}

\subsection{Example: Newton-Cartan non-relativistic geometry from null duality} 
\label{DFTNewtonCartan}

As well as the Gomis-Ooguri limit, the Morand-Park classification can be used to describe various non-relativistic backgrounds. 
Here we will discuss an example which did not in fact appear in \cite{Morand:2017fnv}, namely a version of non-relativistic Newton-Cartan geometry which can be obtained from a Lorentzian geometry by a null T-duality. 
This was used in \cite{Harmark:2017rpg, Harmark:2018cdl} to obtain the Polyakov action for a string in the Newton-Cartan geometry (see also the description of T-duality in the ``string Newton-Cartan'' background of \cite{Bergshoeff:2018yvt}, which showed that a non-relativistic string was T-dual to a Lorentzian background with a null isometry).
This procedure naturally lives in the doubled formalism.
Note that this will therefore again be an example of a $(1,1)$ non-Riemannian generalised metric (whereas the non-relativistic examples in \cite{Morand:2017fnv} had $n\neq \bar n$): one of the advantages of starting with a $(0,0)$ generalised metric and dualising is that we know what happens to the generalised dilaton and can therefore uplift to ExFT later on. 

Here we follow the notation and conventions of \cite{Harmark:2017rpg, Harmark:2018cdl}. We start with the metric for a $d+1$ dimensional Lorentzian spacetime with a null isometry, which can always be put in the form
\be
ds^2 = g_{ij} dx^i dx^j = 2 \tau_\mu dx^\mu ( du - m_\mu dx^\mu ) + h_{\mu\nu} dx^\mu dx^\nu \,,
\label{metric1}
\ee
where $u$ denotes the null direction, and the $d$ dimensional matrix $h_{\mu\nu}$ has rank $d-1$.
The fields $\tau_\mu, m_\nu$ and $h_{\mu\nu}$ together describe a torsional Newton-Cartan geometry and transform under Galilean local symmetries in a particular way that need not concern us here. We also introduce a vector $v^\mu$ and a rank $d-1$ matrix $h^{\mu\nu}$ such that 
\be
h_{\mu\nu} v^\nu = 0 \,,\quad v^\mu \tau_\mu = -1 \,,\quad
h^{\mu\nu} \tau_\nu = 0 \,,\quad h^{\mu\rho} h_{\rho\nu} - v^\mu \tau_\nu = \delta^\mu_\nu \,.
\ee
It is convenient to also define
\be
\bh_{\mu\nu} \equiv h_{\mu\nu} - \tau_\mu m_\nu - \tau_\nu m_\mu \,,\quad
\hv^\mu \equiv v^\mu - h^{\mu\nu} m_{\nu} \,,\quad
\tilde \Phi \equiv - v^\mu m_\mu + \frac{1}{2} h^{\mu\nu} m_\mu m_\nu \,,
\ee
which are invariant under Galilean boosts and rotations, but not under $\mathrm{U}(1)$ gauge transformations of $m_\mu$, $\delta m_\mu = \partial_\mu \lambda$.
In fact, the completeness relation holds with these variables, $h^{\mu\rho} \bh_{\rho\nu} - \hv^\mu \tau_\nu = \delta^\mu_\nu$.
We can then compute the inverses of $g_{ij}$ and $\bh_{\mu\nu}$:
\be
g^{ij} = \begin{pmatrix}
 h^{\mu \nu} & - \hv^\mu \\ 
- \hv^\nu & 2 \tilde \Phi 
\end{pmatrix} \,,\quad
\bh^{\mu\nu} = h^{\mu\nu} - \frac{1}{2\Phi} \hv^\mu \hv^\nu \,.
\ee
We embed the Lorentzian background in the generalised metric $\cH_{MN} = \mathrm{diag}\,( g_{ij}, g^{ij})$. We exchange the direction $u$ for a dual direction $\tilde u$ using the analogue of a Buscher transformation (on the components of the generalised metric, this amounts to swapping the ${}^u$ and ${}_u$ indices). The dual generalised metric is
\be
\cH_{MN} = \begin{pmatrix} 
\bh_{\mu\nu} & 0 & 0 & \tau_\mu \\ 
0 & 2 \tilde \Phi & - \hv^\nu & 0 \\
0 & - \hv^\mu & h^{\mu\nu} & 0 \\
\tau_\nu& 0 & 0 & 0 
\end{pmatrix} \,.
\label{NCGM}
\ee
This does not admit a Riemannian parametrisation. 
Instead, it is again of type $(1,1)$, with:
\be
K_{ij} = \begin{pmatrix} h_{\mu\nu} & 0 \\ 0 & 0 \end{pmatrix} 
\,,\quad
H^{ij} = \begin{pmatrix} h^{\mu\nu} & 0 \\ 0 & 0 \end{pmatrix} 
\,,\quad
B_{ij} = \begin{pmatrix} 0 & -m_\mu \\ m_\nu & 0 \end{pmatrix} \,,\quad
Z_i{}^j =
\begin{pmatrix}
0 & \tau_\mu \\
- v^\nu & 0 
\end{pmatrix}\,,
\ee
where $Z_i{}^j = X_i Y^i - \bar X_i \bar Y^i$ with
\be
X_i = \frac{1}{\sqrt{2}} \begin{pmatrix} \tau_\mu \\ 1 \end{pmatrix} \,,\quad
\bX_i = \frac{1}{\sqrt{2}}\begin{pmatrix} \tau_\mu\\ -1 \end{pmatrix} \,,\quad
Y^i = \frac{1}{\sqrt{2}} \begin{pmatrix} - v^\mu \\  1 \end{pmatrix} \,,\quad
\bY^i = \frac{1}{\sqrt{2}} \begin{pmatrix} - v^\mu \\ - 1 \end{pmatrix} \,.
\ee
Observe that generalised diffeomorphisms include gauge transformations $\delta B_{ij} = 2 \partial_{[i} \lambda_{j]}$, which provide the $\mathrm{U}(1)$ transformations of $m_\mu$ on noting we have $\partial_i = (\partial_\mu, \partial_{\tilde u})$ and $\partial_{\tilde u} = 0$ (i.e. we do not depend on the direction dual to the original isometry direction). This is as expected, as prior to dualising these transformations were part of the diffeomorphism symmetry of the original metric.

The other DFT field which is present is the generalised dilaton, which is invariant under $\mathrm{O}(D,D)$ and so given by $e^{-2\gendil} = \sqrt{|\det g|}$, where 
\be
\det g  = 
- \frac{1}{(d-1)!} \tau_\mu \tau_\nu \eta^{\mu\mu_1\dots\mu_d}\eta^{\nu\nu_1\dots\nu_d} h_{\mu_1\nu_1} \dots h_{\mu_d \nu_d}\,,
\label{detgNC}
\ee
which if we define the $d \times d$ matrix $e_\mu{}^A \equiv ( \tau_\mu, h_\mu{}^a)$, with a $d \times (d-1)$ vielbein $h_\mu{}^a$ such that $h_{\mu\nu} = h_\mu{}^{a} \delta_{ab} h_\nu{}^b$ is just $\det g = - (\det e)^2$.\footnote{Alternatively we can evaluate \eqref{detgNC} by replacing $h_{\mu\nu}$ with $\bh_{\mu\nu}$ so that $\det g = - \det \bh \,\bh^{\mu\nu} \tau_\mu \tau_\nu$, hence
\be
\det g = 
\frac{\det \bh}{2\Phi} = -\frac{1}{ \frac{1}{(d-1)!} \eta_{\mu_1 \dots \mu_d} \eta_{\nu_1 \dots \nu_d} v^{\mu_1} v^{\nu_1} h^{\mu_2 \nu_2} \dots h^{\mu_d \nu_d} }\,.
\ee}
This provides a measure factor for the Newton-Cartan geometry.

It is straightforward to check that inserting the above non-Riemannian generalised metric \eqref{NCGM} in a doubled sigma model and integrating out the dual coordinates reproduces the string action of \cite{Harmark:2018cdl} (one can use for instance the general result of \cite{Morand:2017fnv}).

\section{Riemannian backgrounds and exotic supergravities in $\Gfour$ ExFT} 

We will now focus on the $\Gfour$ ExFT \cite{Berman:2011cg, Blair:2013gqa, Musaev:2015ces}, a good testing ground as it is simple enough to allow one to realise various constructions very explicitly, and simultaneously complex enough to be interesting. 
Already at the level of Riemannian parametrisations, the $\Gfour$ ExFT describes not only the conventional 10- and 11-dimensional supergravities, but exotic variants \cite{Blair:2013gqa}, with all information about the nature of the spacetime theory encoded in the generalised metric via the choice of parametrisation.
We should however note that though these exotic variants appear to give valid parametrisations of the ExFT variables, their role in the full quantum string and M-theory is less clear as they involve spacetimes of non-Minkowskian signatures, and they are not expected to exist as the low energy limits of fully fledged variants of string and M-theory, though they may still appear as complex saddle points in the path integral. 

\subsubsection*{Spacetime decompositions}

In general, in order to match exceptional field theory with standard supergravity, it is convenient to start with an intelligent decomposition of the fields of the latter. 
For instance, the 11- or 10-dimensional Einstein frame metric $\hat g_{\hmu\hnu}$ can be decomposed in the following manner (corresponding to a partial fixing of Lorentz symmetry): splitting the 11- or 10-dimensional index $\hmu = (\mu, i)$, where $\mu$ is an $n$-dimensional index, let
\be
\hat g_{\hmu \hnu} = 
\begin{pmatrix} |\phi|^{\omega} g_{\mu \nu} + A_\mu{}^ k  A_\nu{}^ l \phi_{ k  l} & A_\mu{}^ k  \phi_{ k  j} \\
A_\nu{}^ k  \phi_{ k  i} & \phi_{ i j} 
\end{pmatrix} \,,
\label{eq:GKK}
\ee
where $\omega$ is the intrinsic weight appearing in the generalised Lie derivative (listed in Table \ref{GHR}).
For $\Gfour$, $\omega=-1/5$.
The ExFT formalism will work regardless of the signatures of the blocks $g_{\mu\nu}$ and $\phi_{ij}$.
We will denote the signature of metrics by $(t,s)$. 
Let $\phi_{i  j }$ be a $d$-dimensional metric with signature $(t,s)$, so that $\phi \equiv \det \phi = (-1)^t | \phi |$. 
Define $\epsilon_{i_1\dots i_d} = |\phi|^{1/2} \eta_{i_1\dots i_d}$, $\epsilon^{i_1\dots i_d} = |\phi|^{-1/2} \eta^{i_1\dots i_d}$ with both $\eta^{1\dots d} = \eta_{1 \dots d} = +1$. Then we have $\epsilon^{i_1\dots i_d} = (-1)^t \phi^{i_1 i_1^\prime} \dots \phi^{ i_d i_d^\prime} \epsilon_{i_1^\prime \dots i_d^\prime}$ and there are no extra signs in the contractions between $\epsilon$ with indices up and those with indices down. 

As well as the metric, it can be convenient to redefine the components of the gauge fields which carry the external $\mu,\nu$ indices, making use of the field $A_\mu{}^i$. 
The details are not important in the present paper. The reader can consult the appendices of \cite{Blair:2018lbh} for details adapted to the $\Gfour$ case.

\subsection{The $\mathrm{SL}(5)$ ExFT} 

For $\Gfour$, the representation $R_1$ is the antisymmetric 10-dimensional representation; we will write an $R_1$ index $M$ as an antisymmetic pair of five-dimensional indices $a,b$, so that $V^M \equiv V^{ab} = - V^{ba}$. We will contract indices with a factor of $1/2$, $V^M W_M \equiv \frac{1}{2} V^{ab} W_{ab}$, meaning that $\delta^M{}_N =2 \delta^{[ab]}_{cd} = \delta^a_c \delta^b_d - \delta^b_c \delta^a_d$.
The generalised Lie derivative is defined by giving the Y-tensor, which is $Y^{MN}{}_{KL} = \eta^{aa^\prime bb^\prime e} \eta_{cc^\prime dd^\prime e}$, and the section condition is $\eta^{abcde} \partial_{bc} \partial_{de} = 0 $. 

The generalised metric, $\gM_{MN}$, carries a pair of symmetric $R_1$ indices. 
We can also define a ``little'' generalised metric in the fundamental five-dimensional representation, such that
\be
\gM_{ab,cd} = \pm ( m_{ac} m_{bd} - m_{ad} m_{bc} ) \, ,
\label{bigMlittlem}
\ee
where the overall sign is needed to describe exceptional field theory in the case where the $Y^M$ coordinates include timelike directions. The little metric is constrained to have unit determinant, $\det m_{ab} = 1$.
Note that it is immediate from this decomposition that $\epsilon^{abcde} \gM_{ab,cd} = 0$ and hence $Y^{MN}{}_{PQ} \gM_{MN} = 0$, so that referring to the projector trace $P_{MN}{}^{MN}$ in \eqref{traceP} we find that $\gM_{ab,cd}$ has 14 components, corresponding to the coset $\Gfour/\Hfour$ (or $\Gfour/\mathrm{SO}(2,3)$).
The situation with the sign choice in \eqref{bigMlittlem}, meanwhile, is a little subtle. We choose to fix the sign differently in different parametrisations, such that the ``generalised line element'' 
\be
g_{\mu\nu} dX^\mu dX^\nu + \gM_{MN} ( dY^M + \Aa_\mu dX^\mu)( dY^N + \Aa_\nu dX^\nu ) 
\label{gle}
\ee
when written out in terms of the spacetime metric, $\hat g_{\hmu\hnu}$ (as in \eqref{eq:GKK}), and spacetime coordinates, $\hat X^{\hmu} = (X^\mu, Y^i)$,
always equals
\be
|\phi|^{-\omega} \hat g_{\hmu \hnu} dX^{\hmu} dX^{\hnu} + \dots 
\ee
where the ellipsis denotes terms involving dual coordinates. Pullbacks of the expression \eqref{gle} are used to construct particle and string actions with target space the extended geometry of ExFT, and the relative sign between the two terms is fixed by the appropriate notion of gauge covariance under the ExFT gauge symmetries \cite{Blair:2017gwn, Arvanitakis:2018hfn}.
As it is $\gM_{MN}$ that appears in \eqref{gle}, we stress that it is the parametrisation of this version of the generalised metric which must be considered fundamental, though we will almost always write down explicit expressions using the more compact notation of the little metric $m_{ab}$.  (Note we can also express $m_{ab}$ via $m_{ab} = \frac{1}{6} \eta_{aMN} \eta_{bPQ} \gM^{MP} \gM^{NQ}$.)

The gauge fields of the $\Gfour$ ExFT appearing in the action are a one-form $\Aa_\mu{}^M$, two-form, $\Ab_{\mu\nu a}$ with field strength $\mathcal{H}_{\mu\nu\rho a}$, and three-form, $\Ac_{\mu\nu\rho}{}^a$, whose field strength $\mathcal{J}_{\mu\nu\rho\sigma}{}^a$ appears in the Chern-Simons term but does not have a kinetic term. The equation of motion for $\Ac_{\mu\nu\rho}{}^a$ accordingly amounts to a duality relation relating it to the degrees of freedom in the other gauge fields. 
The action is defined by
\be
\begin{split}
S  = \int \textrm{d}^{7}X \textrm{d}^{10} Y   \sqrt{|g|} &\,\Bigg(
\hat R[g]  
+ \frac{1}{12 } g^{\mu \nu} D_\mu \mathcal{M}_{MN} D_\nu \mathcal{M}^{MN} - V( \mathcal{M},g ) + \frac{1}{\sqrt{|g|}} \mathcal{L}_{CS} 
\\ & \qquad\qquad
- \frac{1}{4} e g^{\mu\rho} g^{\nu \sigma} \gM_{MN} \mathcal{F}_{\mu\nu}{}^M \mathcal{F}_{\rho\sigma}{}^N
- \frac{1}{2} m^{ab} \mathcal{H}_{\mu\nu\rho a} \mathcal{H}^{\mu\nu\rho}{}_{b}
\Bigg) 
\end{split}
\label{actionSL5}
\ee
where
\be
\begin{split}
- V(\gM, g) & = 
\frac{1}{12} \gM^{MN} \partial_M \gM^{KL} \partial_N \gM_{KL} - \frac{1}{2} \gM^{MN} \partial_M \gM^{KL} \partial_K \gM_{LN} 
+ \frac{1}{2} \partial_M \gM^{MN} \partial_N \ln |g|
\\ & \qquad+ \frac{1}{4} \gM^{MN} \left( \partial_M g_{\mu\nu} \partial_N g^{\mu\nu} + \partial_M \ln |g| \partial_N \ln |g| \right)  
\\
& = \pm\Bigg(
 \frac{1}{8}m^{a c} m^{b d} \partial_{a b} m_{ e f} \partial_{c d}m^{e f}+\frac{1}{2}m^{a c} m^{b d} \partial_{a b}m^{e f} \partial_{e c} m_{ d f}+\frac{1}{2}\partial_{a b}m^{a c} \partial_{c d}m^{b d}
 \\ &  \qquad\qquad  + \frac{1}{2} m^{ac}\partial_{ab} m^{b d} \partial_{cd} \ln |g| + \frac{1}{8}m^{a c} m^{b d} ( \partial_{a b} g^{\mu \nu} \partial_{c d} g_{\mu \nu} + \partial_{a b} \ln |g| \partial_{c d} \ln |g| )
 \Bigg) 
\end{split} 
\ee
and the Chern-Simons term is described in \cite{Musaev:2015ces}.

\subsection{M-theory parametrisations} 
\label{5params}

The M-theory solution of the section condition is based on splitting $a=(i,5)$, where $i$ is a four-dimensional index, and choosing the physical coordinates to be $Y^i \equiv Y^{i5}$ and the dual coordinates to be $Y^{ij}$, with the section condition solution then provided by $\partial_{i} \neq 0$, $\partial_{ij} = 0$. 
Generalised diffeomorphisms are generated by $\Lambda^{ab} = ( \Lambda^{i5} , \Lambda^{ij} )$. The vector $\Lambda^i$ is then found to generate four-dimensional diffeomorphisms, while $\Lambda^{ij} = \frac{1}{2} \eta^{ijkl} \lambda_{kl}$ produces gauge transformations of the three-form.
This allows us to parametrise the generalised metric in terms of the internal spacetime metric, $\phi_{ij}$, and the internal components of the three-form, $C_{ijk}$. 
It is convenient to turn $C_{ijk}$ into a vector by defining $v^ i \equiv \frac{1}{3!} \epsilon^{ i j k  l} C_{ j k  l}$.
Then we have:
\be
m_{ab} = 
\begin{pmatrix}
\lambda |\phi|^{-2/5} \phi_{ i  j} & -\lambda |\phi|^{1/10} v_ i \\
-\lambda |\phi|^{1/10}  v_ j & |\phi|^{3/5} ( (-1)^t + \lambda v^ k  v_ k ) 
\end{pmatrix} 
\,.
\label{mparam}
\ee
This parametrisation incorporates two sign factors.
The first of these is $(-1)^t$, which depends on the number of timelike directions $t$ in $\phi_{ij}$. 
This appears in order that the generalised metric parametrise the correct coset $\mathrm{SL}(5) / \mathrm{SO}(2,3)$ rather than $\mathrm{SL}(5) / \mathrm{SO}(5)$, and ensures that the determinant is $+1$. Such timelike variants of the classic $G/H$ cosets were analysed in \cite{Hull:1998br}.
The second sign factor is denoted by $\lambda$, and controls the sign of the kinetic term of the three-form, providing an ExFT parametrisation for exotic variants of 11-dimensional supergravity related to timelike dualities \cite{Hull:1998vg,Hull:1998ym}. See also \cite{Blair:2013gqa} for an earlier discussion of such parametrisations in the $\Gfour$ ExFT. 
The parametrisation of the big generalised metric that we use corresponds to
\be
\gM_{ab,cd} = \lambda (-1)^t ( m_{ac} m_{bd} - m_{ad} m_{bc} ) .
\ee
Studying the gauge transformations of the ExFT gauge fields in this solution of the section condition, we find that the obvious components of the 11-dimensional three-form can be identified with certain components of the ExFT gauge fields, schematically $\Aa_{\mu}{}^{ij} = \frac{1}{2} \eta^{ijkl} C_{\mu kl}$, $\Ab_{\mu \nu i} = C_{\mu\nu i}$, $\Ac_{\mu\nu\rho} = C_{\mu\nu\rho}$ (see the appendices of \cite{Blair:2018lbh} for more precise relationships). Apart from the obvious identification $\Aa_\mu{}^i = A_\mu{}^i$, the other components of the gauge fields are related to the dual 11-dimensional six-form, and can be eliminated from the ExFT action using duality relations. 
As a result, one finds by explicit calculation that the ExFT action is equivalent to that of 11-dimensional supergravity:
\be
\begin{split} 
	S & = \int d^{11}X \sqrt{|\hat g|} \left( R(\hat g) - \lambda \frac{1}{48} F^{\hmu \hnu \hrho\hsigma} F_{\hmu \hnu \hrho\hsigma} + \frac{1}{\sqrt{|\hat g|}} \mathcal{L}_{\text{CS}} \right) \,.
\end{split} 
\ee
In general we see that $\lambda = +1$ corresponds to the usual relative sign between the Ricci scalar and $F^2$ term, while $\lambda = -1$ flips the sign of the $F^2$ term. 
The latter variant of supergravity can be thought of as the low energy effective action of an exotic M-theory, called M${}^-$ theory, of signature $(2,9)$ and containing M2 branes whose worldvolume has Euclidean signature \cite{Hull:1998vg,Hull:1998ym, Dijkgraaf:2016lym}.

\begin{figure}[ht]
\centering
\begin{tikzpicture}[scale=0.9,black]
\tikzstyle{theory}=[align=center,draw=black!0,text width=2cm, font=\scriptsize]
\draw [blue!40, rounded corners = 0.5cm, thick, fill=blue!0] (-8.2,0.6) rectangle (-0.8,6.4);
\draw [color =black] (-5.5,0) node [align=left] { \scriptsize  DFT$^+$ \scriptsize \cite{Hohm:2011dv} };
\draw [ green!40, rounded corners = 0.5cm, thick,  fill=green!0] (0.8,0.6) rectangle (8.2,6.4);
\draw [color=black] (6.75,0) node [align=left] {\scriptsize  DFT$^-$   \scriptsize \cite{Blair:2016xnn}};
\begin{scope}[xshift=-1cm]
\draw (-3.5,5.5) node[theory] {\scriptsize  IIA$^{++}$ \scriptsize (IIA) \\ \scriptsize (1,9) };
\draw (-5.5,3.5) node[theory] {\scriptsize  IIB$^{++}$ \scriptsize (IIB) \\\scriptsize  (1,9) };
\draw (-1.5,3.5) node[theory] {\scriptsize  IIB$^{+-}$ \scriptsize (IIB$^*$) \\\scriptsize  (1,9) };
\draw (-3.5,1.5) node[theory] {\scriptsize  IIA$^{+-}$ \scriptsize (IIA$^*$) \\\scriptsize  (1,9) };
\draw[<->] (-4.75,5.5) -- (-5.5,4);
\draw[<->] (-4.75,1.5) -- (-5.5,3);
\draw[<->] (-2.25,5.5) -- (-1.5,4);
\draw[<->] (-2.25,1.5) -- (-1.5,3);
\draw (-5.25,4.85) node {\scriptsize $x$};
\draw (-1.65,2.25) node {\scriptsize $x$};
\draw (-5.4,2.25) node {\scriptsize $t$};
\draw (-1.75,4.85) node {\scriptsize $t$};
\end{scope}
\begin{scope}[xshift=1cm]
\draw (1.5,3.5) node[theory] {\scriptsize IIB$^{-+}$ \scriptsize (IIB$^\prime$) \\ \scriptsize (1,9) };
\draw (3.5,5.5) node[theory] {\scriptsize IIA$^{-+}$ \scriptsize (IIA$_E$) \\ \scriptsize (0,10)};
\draw (3.5,1.5) node[theory] {\scriptsize IIA$^{--}$ \\ \scriptsize (2,8) };
\draw (5.5,3.5) node[theory] {\scriptsize IIB$^{--}$  \\ \scriptsize (3,7) };
\draw (4.75,4.55) node {$\dots$};
\draw[<->] (4.75,1.5) -- (5.5,3);
\draw[<->] (2.25,5.5) -- (1.5,4);
\draw[<->] (2.25,1.5) -- (1.5,3);
\draw (2,5.4 ) node {\scriptsize $x$};
\draw (1.5 ,4.3 ) node {\scriptsize $t$};
\draw ( 1.5,2.7 ) node {\scriptsize $x$};
\draw (5.6,2.7) node {\scriptsize $t$};
\draw ( 2,1.6 ) node {\scriptsize $t$};
\draw (5,1.6) node {\scriptsize $x$};
\end{scope}
\draw (0,7) node[align=center] {\small M$^+$ \footnotesize (M) \\ \scriptsize (1,10) };
\draw (0,0) node[align=center] {\small M$^-$ \footnotesize (M$^*$) \\ \scriptsize (2,9) };
\draw[red,thick,->] (-0.9,7) --(-3.4,6.1);
\draw[red,thick,->] (0.9,7) --(3.4,6.1);
\draw[red,thick,->] (-0.9,0) --(-3.4,.9);
\draw[red,thick,->] (0.9,0) --(3.4,.9);
\draw[red] ( -2,0.15 ) node {\scriptsize $t$};
\draw[red] (2,0.15) node {\scriptsize $x$};
\draw[red] ( -2,6.8 ) node {\scriptsize $x$};
\draw[red] (2,6.8) node {\scriptsize $t$};
\draw [<->,densely dashed] (1.2,3.5) -- (-1.2,3.5);
\draw (0,3.8) node {\scriptsize S};
\end{tikzpicture} 
\caption{The exotic duality web. Red arrows denote timelike or spacelike reductions from 11 to 10 dimensions. Black arrows denote T-dualities. The dashed arrow in the centre denotes S-duality. All these theories are described by choosing different parametrisations of exceptional field theory. 
The superscript IIA/B${}^{\pm \pm}$ denotes whether, firstly, fundamental strings and, secondly, D-branes have Lorentzian or Euclidean worldvolumes, and hence determines which gauge fields have wrong sign kinetic terms. Similarly M${}^\pm$ denotes whether M2 branes have Lorentzian or Euclidean worldvolumes. There are additional versions of these theories with more exotic signatures. The alternate names in brackets are those used originally by Hull \cite{Hull:1998vg,Hull:1998ym}, while the plus minus notation and form of the diagram is taken from \cite{Dijkgraaf:2016lym}.
}
\label{exoticdiagram}
\end{figure}
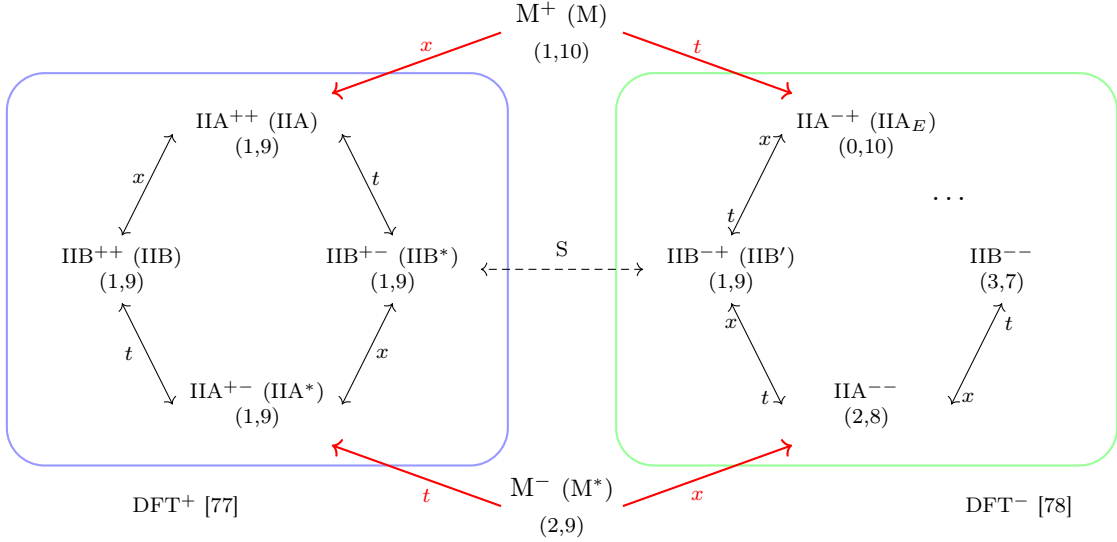

We can summarise some of the sign choices appearing in the little generalised metric \eqref{mparam}, with reference to figure \ref{exoticdiagram}:
\begin{itemize}
	\item The signature of $\phi_{ij}$ is $(0,4)$ and $\lambda = +1$ so that the signature of $m_{ab}$ is $(0,5)$, and if the external metric has signature $(1,6)$ this describes the usual 11-dimensional SUGRA. 
	\item The signature of $\phi_{ij}$ is $(1,3)$ and $\lambda = +1$ so that the signature of $m_{ab}$ is $(2,3)$, and if the external metric has signature $(0,7)$ this describes the usual 11-dimensional SUGRA. 
	\item The signature of $\phi_{ij}$ is $(2,2)$, and $\lambda = -1$ so that the signature of $m_{ab}$ is $(2,3)$, and if the external metric has signature $(0,7)$ this describes the unusual 11-dimensional SUGRA with signature $(2,9)$ and wrong sign kinetic term, the low energy limit of the $M^*$ theory (see diagram \ref{exoticdiagram}).
	\item Other choices can correspond to ExFT descriptions of other exotic variants of M-theory.
\end{itemize}

\subsection{IIB parametrisations}

For the IIB solution of the section condition we split $a=(i,\alpha)$ where $i$ a three-dimensional index, and $\alpha$ is a two-dimensional index associated to the unbroken $\mathrm{SL}(2)$ S-duality symmetry of IIB. 
The physical coordinates are then the three coordinates $Y^{ij}$. 
It can be convenient to view the $i$ index as being naturally down, i.e.\ $Y^M = (Y_{ij}, Y_i{}^\alpha, Y^{\alpha \beta})$, such that the physical coordinates can be defined to have the usual index position via $Y^i = \eta^{ijk} Y_{jk}$.

The generalised diffeomorphism parameter $\Lambda^{ab} = ( \eta_{ijk} \Lambda^k, \Lambda_i{}^\alpha, \Lambda^{\alpha \beta})$ now produces three-dimensional diffeomorphisms generated by $\Lambda^i$, gauge transformations $\Lambda_i{}^\alpha$ of the two-form doublet, and gauge transformations $\Lambda^{\alpha \beta} \equiv \varepsilon^{\alpha \beta} \frac{1}{3!} \eta^{ijk} \lambda_{ijk}$ of the four-form singlet. 

The generalised metric can be parametrised in terms of the internal metric, $\phi_{ij}$, the two two-forms $(C_{ij}, B_{ij}) = C_{ij}{}^\alpha$ (which we again write as vectors, $v^{i \alpha} \equiv \frac{1}{2} \epsilon^{ijk} C_{jk}{}^\alpha$), and a two-by-two matrix, $\cH_{\alpha \beta}$, containing the dilaton $\Phi$ and RR zero-form $C_0$. 
We write 
\begin{align}
m_{ab} & =
	\begin{pmatrix}
|\phi|^{3/5} ( (-1)^t \sigma_F \sigma_D \phi^{ij} + \cH_{\gamma \delta} v^{i \gamma} v^{j \delta} ) &  |\phi|^{1/10} \cH_{\alpha \gamma} v^{i \gamma} \\
  |\phi|^{1/10}  \cH_{\beta \gamma} v^{j \gamma} & |\phi|^{-2/5} \cH_{\alpha \beta}
\end{pmatrix}  \,,\\
\cH_{\alpha \beta} & = 
\sigma_F e^\Phi \begin{pmatrix} 
1 & C_0 \\ C_0 & \sigma_F \sigma_D e^{-2\Phi} + C_0^2 
\end{pmatrix} \,.
\label{mIIBparam}
\end{align}
Again, we allow for a general distribution of sign factors when the coset is $\mathrm{SL}(5) / \mathrm{SO}(2,3)$.
Here the signs $\sigma_i = \pm$ dictate whether the parametrisation corresponds to a set of variants of type IIB, denoted IIB${}^{\sigma_F \sigma_D}$, where IIB${}^{++}$ is the standard IIB, IIB${}^{+-}$ is obtained by a timelike T-dualisation of type IIA, IIB${}^{-+}$ is the S-dual of IIB${}^{+-}$ and is a theory where the fundamental strings have Euclidean worldsheet, and IIB${}^{--}$ is obtained by further T-dualities \cite{Hull:1998vg,Hull:1998ym, Dijkgraaf:2016lym}.
The subscript on $\sigma_F$ means that the sign corresponds to the F1 having Lorentzian/Euclidean worldvolume, while that on $\sigma_D$ means that the sign corresponds to D-branes having Lorentzian/Euclidean worldsheets. 
In this case, the parametrisation of the big generalised metric that we use corresponds to
\be
\gM_{ab,cd} = (-1)^t ( m_{ac} m_{bd} - m_{ad} m_{bc} ) .
\ee
We also identify the gauge fields such that (schematically) $\Aa_{\mu ij} = \eta_{ijk} A_\mu{}^k$, $\Aa_{\mu i}{}^\alpha = (C_{\mu i} , B_{\mu i} )$, $\Aa_{\mu}{}^{\alpha \beta} = \varepsilon^{\alpha \beta} \frac{1}{3!} \eta^{ijk} C_{\mu ijk}$ and similarly for the higher rank fields. 
Then the $\Gfour$ ExFT dynamics are equivalent to those following from the type pseudo-IIB action\footnote{The self-duality of the five-form field strength is imposed in the $\Gfour$ ExFT by the duality relations corresponding to the equation of motion of the gauge field $\Ac_{\mu\nu\rho}{}^a$.} of the form
\be
\begin{split} 
S = \int d^{10}X \,\sqrt{|\hat g|} &\Big(
R(\hat g) + \frac{1}{4} \hat g^{\hmu \hnu} \partial_{\hmu} \cH_{\alpha \beta} \partial_{\hnu} \cH^{\alpha \beta} - \frac{1}{12} \sigma_D \sigma_F \cH_{\alpha \beta} {F}_{\hmu \hnu \hrho}{}^\alpha F^{\hmu \hnu \hrho \beta}
\\ & \qquad   - \frac{1}{4\cdot 5!} \sigma_D \sigma_F F_{\hmu_1 \dots \hmu_5} F^{\hmu_1 \dots \hmu_5} 
+ \frac{1}{\sqrt{|\hat g|}} \mathcal{L}_{\text{CS}}
\Big),
\end{split}
\ee
which matches the Einstein frame action exactly for the type IIB${}^{\sigma_F \sigma_D}$ supergravities \cite{Dijkgraaf:2016lym}.
We see that the choice of signs $\sigma_F, \sigma_D$ will determines which kinetic terms come with the wrong sign. When $\sigma_F = -1$, the NSNS $B$-field does, while when $\sigma_D=-1$ the RR two-form does.

We can summarise some of the sign choices appearing in the little generalised metric \eqref{mIIBparam}, with reference to figure \ref{exoticdiagram}:
\begin{itemize}
	\item The signature of $\phi_{ij}$ is $(0,3)$, then we can describe either usual IIB${}^{++}$, in the $\Gfour/\mathrm{SO}(5)$ description, with the external metric of signature $(1,6)$, or also IIB${}^{--}$, in the $\Gfour/\mathrm{SO}(2,3)$ description.
	\item The signature of $\phi_{ij}$ is $(1,2)$, then we can describe IIB${}^{++}$ (with Euclidean external metric) as the $(-1)^t$ factor flips the signature of the upper three-by-three block: this way we describe the usual Lorentzian supergravity (in \cite{Blair:2013gqa} this was viewed as working with a mostly minus signature in the spacetime picture). We can also describe IIB${}^{+-}$ or IIB${}^{-+}$.
	\item Other choices can correspond to ExFT descriptions of other exotic variants of IIB supergravity.
\end{itemize}

\section{Non-Riemannian backgrounds in $\Gfour$ ExFT}
\label{5NonRie}

We will now generate and describe non-Riemannian parametrisations of the $\Gfour$ theory. We will first play with the same simple examples as worked in the DFT case: namely, a Gomis-Ooguri scaling limit, and a U-duality of the worldvolume directions of an M2 brane solution. 
We will then demonstrate how to think more systematically about such parametrisations in this ExFT.

\subsection{Examples: Gomis-Ooguri and timelike U-duality} 

\subsubsection*{Gomis-Ooguri}

Consider the flat background
\be
ds^2 = G^{2/3} \eta_{\alpha \beta} dz^\alpha dz^\beta + G^{-1/3} d\vec{x}_8{}^2
\,,\quad
C_{t12} = G - \mu \,,
\ee
where $\eta_{\alpha \beta} = \mathrm{diag}\,(-1,1,1)$ and $G$ and $\mu$ are tunable constants: we will take the $G \rightarrow \infty$ limit keeping $\mu$ fixed.
The relative scaling between the worldvolume coordinates $z^\alpha = (t,z^1,z^2)$ and the transverse coordinates $\vec{x}_8$ is the same as in the M2 example considered in \cite{Gomis:2000bd}, however we need to rescale the overall metric by a factor of $G^{2/3}$ in order to obtain a finite generalised metric in the limit. 
We pick ExFT physical coordinates $Y^{i5} = (z^\alpha,w)$ where $w$ denotes any one of the $\vec{x}_8$ directions. We take $\eta^{t12w} = +1$. 
The only non-vanishing ExFT fields are the external metric and generalised metric (using the parametrisation \eqref{mparam} with, here, $\lambda = t = + 1$):
\be
g_{\mu\nu} = \delta_{\mu\nu}\,,\quad
m_{ab} = \begin{pmatrix}
\eta_{\alpha \beta} & 0 & 0 \\
0 & G^{-1} &  1-\mu G^{-1}  \\
 0 & 1 - \mu G^{-1}  & -2\mu + \mu^2 G^{-1} 
\end{pmatrix} \,.
\ee
The $G \rightarrow \infty$ limit is well-defined and non-singular at the level of the generalised metric, leading to
\be
m_{ab} = \begin{pmatrix}
\eta_{\alpha \beta} & 0 & 0 \\
0 & 0 & 1  \\
 0 & 1  & -2\mu  
\end{pmatrix} \,.
\label{exGOm}
\ee
However, this does not admit a Riemannian parametrisation: the upper left 4 by 4 block, which should be proportional to the spacetime metric, is degenerate.
We have therefore very easily generated an example of a non-Riemannian parametrisation in exceptional field theory.

\subsubsection*{Non-Riemannian geometry from U and S duality}

Our next trick will involve dualising supergravity solutions.
We will apply the following U-duality transformation:
\be
U^a{}_b = \begin{pmatrix}
\delta^i{}_j  - n^i \bar n_j & n^i \\
- \bar n_j & 0 
\end{pmatrix} 
\label{UBuscher}
\ee
where $n^i \bar n_i = 1$. If the physical directions are indexed by $i=1,2,3,4$ and we want to do a U-duality acting on the $1,2,3$ directions, then we let $n^4 = 1 = \bar n_4$. This U-duality then reduces (after reduction on the $3$ direction, say) to a pair of Buscher T-dualities acting in the $1,2$ directions plus an interchange of the $x^1$ and $-x^2$ directions.
The generalised metric transforms to $\tilde m = U^{-T} m U^{-1}$.

\subsubsection*{U-duality between M2 and non-Riemannian background}

The M2 solution, which in fact inspired the form of the Gomis-Ooguri background considered above, is
\be
\begin{split}
ds^2 = H^{-2/3} \eta_{\alpha \beta} dz^\alpha dz^\beta + H^{1/3} d \vec{x}_8{}^2 \,,\quad
C_{t12}  = H^{-1} + c \,,\quad
H = 1 + \frac{h}{|\vec{x}_8|^6}\,.
\end{split} 
\ee
One would normally have $c=-1$, such that the three-form vanishes at infinity, however we leave this constant general (corresponding to a large gauge transformation at infinity). 
We again pick ExFT physical coordinates $Y^{i5} = (z^\alpha,w)$ with $z^\alpha = (t,z^1,z^2)$ and  $w$ denoting one of the $\vec{x}_8$ directions. 
The ExFT embedding of this solution (again, with $\lambda = t = +1$) is
\be
g_{\mu\nu}  = \delta_{\mu\nu} \,,\quad
m_{ab} = \begin{pmatrix} 
\eta_{\alpha \beta} & 0 & 0 \\
0 & H & 1 + H c \\
0 &  1 + H c & 2c + Hc^2 
\end{pmatrix} \,.
\label{M2inm}
\ee
We dualise on the (isometric) $(t,z^1,z^2)$ directions. The $\mathrm{SL}(5)$ transformation \eqref{UBuscher} and transformed generalised metric are:
\be
U^a{}_b = \begin{pmatrix}
\delta^\alpha_\beta & 0 & 0 \\ 
 0 & 0 & 1 \\
 0 & -1 & 0
\end{pmatrix} \,,
\quad \tilde m_{ab}  = \begin{pmatrix} 
\eta_{\alpha \beta} & 0 & 0 \\
0  &  2c + Hc^2  & -(1+Hc)\\
0 & -(1+Hc) & H
\end{pmatrix} \,.
\label{mafterU}
\ee
The spacetime parametrisation that one should use for $\tilde m_{ab}$ depends on the value of $\tilde H \equiv - 2c - Hc^2$.

\begin{itemize} 

\item If $\tilde H < 0$, then the signature of the upper left 4 by 4 block is still $(1,3)$ so this admits the conventional geometric parametrisation.
We find
\be
\begin{split}
ds^2 = (|\tilde H|)^{-2/3} ( - dt^2 + (dz^1)^2 + (dz^2)^2 ) + (|\tilde H|)^{1/3} d \vec{x}_8{}^2 \,,\quad
C_{t12} = - \tilde H^{-1} - c^{-1} \,.
\end{split} 
\label{negM2in}
\ee

\item If $\tilde H > 0$, then the signature of the upper left 4 by 4 block is now $(2,2)$. It seems we have a choice of whether to use the parametrisation with $\lambda=1$ or $\lambda = -1$. The latter preserves the $S\mathrm{O}(8)$ invariance of the transverse directions, giving
\be
\begin{split}
ds^2  = (\tilde H)^{-2/3} ( dt^2 - (dz^1)^2 - (dz^2)^2 ) + (\tilde H)^{1/3} d \vec{x}_8{}^2 \,,\quad
C_{t12}  = - \tilde H^{-1} - c^{-1} \,.
\end{split} 
\label{negM2out}
\ee

\end{itemize} 

For $c=-1$ we have $\tilde H = 2 - H = 1 - h/r^6$. Then the solution \eqref{negM2in} and \eqref{negM2out} describes a negative M2 in the exotic M${}^-_{2,9}$ theory with two timelike directions. In the region outside the brane where $\tilde H > 0$ and the spacetime is described by \eqref{negM2out}. Passing through the singularity at $\tilde H = 0$, the interior ($\tilde H < 0$) configuration \eqref{negM2in} then has as usual flipped signature in the worldvolume directions (so in fact is described by M-theory with conventional signature).\footnote{In \cite{Malek:2013sp} it is argued that the change of signature of spacetime should instead be viewed as a breakdown in the gauge fixing of the generalised vielbein, implying that it is impossible to everywhere describe the background in terms of a metric and a three-form. Instead, owing to the singularity at $\tilde H = 0$, we could introduce a parametrisation involving a dual metric and a trivector which will be globally defined on the dual geometry. Here we instead adopt the perspective that though the generalised metric is well-defined everywhere, the choice of parametrisation that one makes is discontinuous at $\tilde H = 0$. At the singularity $\tilde H = 0$, the geometry becomes non-Riemannian but on either side we have different Riemannian parametrisations in which spacetime has different signatures. Similar comments should apply to the DFT example considered previously, also.}

The generalised metric is non-Riemannian at $\tilde H =0$ when $c \neq 0$, and everywhere when $c=0$, with
\be
\tilde m_{ab}(\tilde H = 0)  = \begin{pmatrix} 
\eta_{\alpha \beta} & 0 & 0 \\
 0 &  0  & 1\\
 0 & 1 & -2/c
\end{pmatrix} \,,\quad
\tilde m_{ab}(c=0) = \begin{pmatrix} 
\eta_{\alpha \beta} & 0 & 0 \\
 0 &  0  & -1\\
 0 & -1 & H
\end{pmatrix}
\,.
\label{exUm}
\ee
These are evidently non-Riemannian in the same manner as the Gomis-Ooguri example we considered previously.

\subsubsection*{S-duality between D$(-1)$ and non-Riemannian background}

We can also consider the generalised metric \eqref{M2inm} of the embedding of the M2 in the $\mathrm{SL}(5)$ ExFT and this time interpret it in a type IIB solution of the section condition. 
The appropriate parametrisation corresponds to the type IIB${}^{+-}$ theory (related by T-duality on the timelike direction to type IIA). 
We find
\be
e^\Phi = H \,,\quad
C_0 = H^{-1} + c\,,
\ee
while the 10-d Einstein metric is flat. 
This can be interpreted as the solution for a D$(-1)$ (normally this is obtained as a solution of the Euclideanised type IIB, a subtlety we will ignore. It seems natural to view the D$(-1)$ here as having been obtained by timelike T-duality of a D0). We can relate this IIB parametrisation back to the M-theory one: the relationship is evidently given by T-dualising the D$(-1)$ on all three physical coordinates to get a D2 which lifts to the M2 we started off by thinking about.

The U-duality transformation in \eqref{mafterU} now in fact corresponds to the S-duality that inverts the string coupling constant. 
The resulting configuration \eqref{mafterU} can be simply described as:
\be
e^\Phi = - \sigma_F \tilde H
\,,\quad
C_0 = - ( \tilde H^{-1} + c^{-1} )
\,,\quad
\sigma_F = - \mathrm{sgn} \,(\tilde H )
\ee
where $\sigma_F$ is determined by requiring $e^\Phi$ be positive.

Let's consider the two cases of most interest:

\begin{itemize}
\item When $c=-1$, we have $\tilde H = 2 -H$ as before, and the solution can be written everywhere as
\be
e^\Phi = | \tilde H |\,,\quad
C_0 = 1 - \tilde H^{-1}\,.
\ee
For $\tilde H > 0$ we have $\sigma_F = -1$, $\sigma_D= +1$, so that the theory corresponds to IIB${}^{-+}$, while for $\tilde H < 0$, we have $\sigma_F = +1$, $\sigma_D=-1$, so this is IIB${}^{+-}$. This therefore is a negative D$(-1)$ in the IIB${}^{-+}$ theory. Note there is no signature flip (as we are dealing with a worldpoint).

\item When $c=0$, there is no parametrisation in terms of $e^\Phi$ and $C_0$, although the spacetime metric is well-defined. The ``non-Riemannian'' nature of the generalised metric is thus that the parametrisation of the part of the generalised metric, $\cH_{\alpha \beta}$, describing the axio-dilaton of type IIB, is non-standard.

\end{itemize}

\subsection{Non-Riemannian little metrics}

We can begin to understand the presence of non-Riemannian parametrisations in this ExFT by looking anew at the generalised metric.
We continue to use the little metric, and will work in an M-theory solution of the section condition.
Then, a general parametrisation of the little metric is:
\be
m_{ab} = \begin{pmatrix} k_{ij} & \chi_i \\ \chi_j & \tilde \varphi \end{pmatrix} \,,
\label{mgeneral}
\ee
assuming that it is symmetric and subject to the sole constraint that $\det m = 1$, such that
\be
\tilde \varphi \det k - \frac{1}{6} \chi_{i_1} \chi_{j_1} \eta^{i_1 i_2 i_3 i_4} \eta^{j_1 j_2 j_3 j_4} k_{i_2j_2} k_{i_3j_3} k_{i_4j_4}  = 1\,.
\label{mdetconstraint}
\ee
We will now attempt to solve this constraint to find allowed parametrisations of $m_{ab}$.

\subsubsection*{Riemannian solutions} 

First, we suppose that $\det k \neq 0$, and let $k^{ij}$ denote the inverse of $k_{ij}$. It follows from \eqref{mdetconstraint} that 
\be
\tilde \varphi = \frac{1}{\det k} \left( 1 + \det k \, k^{ij} \chi_i \chi_j \right) \,.
\ee
Setting $k_{ij} = \lambda |\phi|^{-2/5} \phi_{ij}$ and $\chi_i = \lambda |\phi|^{1/10} \phi_{ij} v^j$ recovers the usual parametrisation \eqref{mparam}, after noting $\operatorname{det} k = {(-1)}^t {|\phi|}^{-3/5}$, consistent with the transformation properties implied by generalised diffeomorphisms. Note that we can have $\tilde \varphi = 0$ when $\phi_{ij} v^i v^j = - \lambda (-1)^t$.

\subsubsection*{Non-Riemannian solutions} 

Now, we suppose $\det k = 0$. This takes us into the realm of non-Riemannian parametrisations, as evidently we will not be able to identify $k_{ij}$ as being proportional to the four-dimensional spacetime metric anymore. 
Define a vector
\be
u^i \equiv  - \frac{1}{6}  \chi_{j_1} \eta^{i\, i_2 i_3 i_4} \eta^{j_1 j_2 j_3 j_4} k_{i_2j_2} k_{i_3j_3} k_{i_4j_4}\,.
\ee
Then the constraint \eqref{mdetconstraint} is equivalent to $u^i \chi_i = 1$, and we have $k_{ij} u^j = 0$. Note that then $k_{ij}$ has rank 3, and we cannot satisfy \eqref{mdetconstraint} if it has lower rank.
We introduce the almost inverse $\tilde h^{ij}$ such that 
\be
\tilde h^{ik} k_{kj} + u^i \chi_j = \delta^i_j \,,\quad  \tilde h^{ij} \chi_j = 0 \,,\quad k_{ij} u^j = 0 \,.
\label{originalrelations}
\ee
Then, the inverse metric is
\be
m^{ab} = \begin{pmatrix} \tilde h^{ij} - \tilde \varphi u^iu^j & u^i \\ u^j & 0 \end{pmatrix} \,.
\label{minvgeneral} 
\ee
Explicitly, $\tilde h^{ij} =- \frac{1}{2} \eta^{iklm} \eta^{jpqr} \chi_m \chi_r k_{kp} k_{lq}$.

The parametrisations \eqref{mgeneral} and \eqref{minvgeneral} together with the relations \eqref{originalrelations} completely encode the non-Riemannian parametrisation. 
To understand the quantities appearing, let us examine the transformation of the generalised metric under generalised diffeomorphisms \cite{Berman:2011cg}
\be
\delta_\Lambda m_{ab} = \frac{1}{2} \Lambda^{cd} \partial_{cd} m_{ab} - \frac{2}{5} m_{ab} \partial_{cd} \Lambda^{cd} + 2 m_{c(a} \partial_{b)d} \Lambda^{cd} \,.
\ee
In the M-theory solution to the section condition, we let $\Lambda^{ab} = ( \Lambda^{i5} , \Lambda^{ij} )$ such that
\be
\delta_\Lambda k_{ij} = L_\Lambda k_{ij} 
\,,\quad
\delta_\Lambda \chi_i  = L_\Lambda \chi_i 
+ k_{ij} \partial_k \Lambda^{kj} \,,\quad
\delta_\Lambda \tilde\varphi  = L_\Lambda \tilde\varphi 
 - 2 \chi_i \partial_k \Lambda^{ik} \,,\quad
\ee
\be
\delta_\Lambda \tilde h^{ij}  = L_\Lambda \tilde h^{ij} 
- 2 \chi_{k} \partial_l \Lambda^{kl} u^i u^j + 2 u^{(i} \partial_k \Lambda^{j) k} \,,\quad
\delta_\Lambda u^i  = L_\Lambda u^i
\,,\quad
\label{localsymms5nonrie}
\ee
where $L_\Lambda$ denotes the usual Lie derivative with respect to the parameter $\Lambda^{i5}$, with $k_{ij}$, $\chi_i$, $\tilde \varphi$, $u^i$ and $\tilde h^{ij}$ of weight $-4/5$, $1/5$, $6/5$, $-1/5$ and $+4/5$ respectively. Normally one views the transformations $\Lambda^{ij}$ as giving gauge transformations of the 11-dimensional three-form, with $\Lambda^{ij} = \frac{1}{2} \eta^{ijkl} \lambda_{kl}$.
We observe that though $k_{ij}$ and its zero vector $u^i$ are invariant under such transformations, the other quantities are not. 
(However the defining properties \eqref{originalrelations} are preserved such the transformed variables still give a valid non-Riemannian parametrisation.)

To see how we could introduce a three-form into the non-Riemannian parametrisation transforming under the gauge transformations, suppose that we start with the degenerate $k_{ij}$ and its zero vector $u^i$.
If all we require of $\chi_i$ is that it obeys $u^i \chi_i = 1$, there is an ambiguity $\chi \sim \chi_i + k_{ij} b^j$ for arbitrary vector $b^j$.
We can provide a partial fixing of this ambiguity by defining a particular covector $X_i$ such that $u^i X_i = 1$.
Taking this as a reference, we can write
\be
\chi_i = X_i - k_{ij} V^j \,,
\ee
and require that $X_i$ be wholly ``geometric'' in the sense that it is unchanged by the gauge transformations $\Lambda^{ij}$. 
Then, we find that
\be
\delta_\Lambda V^i = L_\Lambda V^i + \partial_{k} \Lambda^{ik} \,,
\ee 
(with weight $1$). Simultaneously, we can redefine the other fields transforming under the gauge transformations according to
\be
\tilde h^{ij} = h^{ij} + 2 u^{(i} V^{j)} + (- 2 V^k X_k + V^k k_{kl} V^l ) u^i u^j \,,\quad
\tilde \varphi = \varphi -  2 V^k X_k + V^k k_{kl} V^l,
\ee
in terms of which $\varphi$, $h^{ij}$ are gauge invariant and we have an alternative version of the parametrisation, with
\be
m_{ab} 
=
\begin{pmatrix} 
k_{ij} & X_i - k_{ik} V^k\\
X_j - k_{jk} V^k & \varphi - 2 V^k X_k + V^k k_{kl} V^l 
\end{pmatrix} \,,\quad
\label{ansatz}
m^{ab} = 
\begin{pmatrix}
h^{ij} - \varphi u^i u^j + 2 u^{(i}V^{j)} & u^i \\
u^j & 0 
\end{pmatrix} \,.
\ee
This can be factorised as $m_{ab} = (  U_V^T  \bar m U_V)_{ab}$
where 
\be
(U_V)^a{}_b = \begin{pmatrix} 
\delta^i_j & -V^i \\
0 & 1
\end{pmatrix} \,,\quad \mathrm{det}\,  U_V  = 1 \,,
\quad 
\bar m_{ab} = 
\begin{pmatrix} 
k_{ij} & X_i \\
X_j & \varphi 
\end{pmatrix}\,.
\ee 
The relations in \eqref{originalrelations} are now:
\be
k_{ij} u^j = 0 \,,\quad h^{ij} X_j = 0 \,,\quad u^i X_i = 1\,,\quad
h^{ik} k_{kj} + u^i X_j = \delta^i_j \,.
\label{defining}
\ee
In this version of the parametrisation, as $V^i$ transforms under three-form gauge transformations as $\delta_\lambda V^i = 3 \eta^{ijkl} \partial_j \lambda_{kl}$, we can think of relating this to three-form as $V^i = \frac{1}{3!} \eta^{ijkl} C_{jkl}$.

This form of the generalised metric \eqref{ansatz} appears to contain the degrees of freedom we might expect for $\Gfour/\mathrm{SO}(2,3)$: in place of a Riemannian metric we instead have the degenerate $k_{ij}$ and $X_i$, along with a three-form (encoded in $V^i$) and the extra scalar $\varphi$, subject to the constraint $u^i X_i =1$. However, given the ambiguity in introducing $V^i$, the situation is quite subtle.
If we insist on using the parametrisation \eqref{ansatz}, then we must note that it is invariant under the shift symmetry:
\be
 X_i \rightarrow X_i + k_{ij} b^j \, ,  \qquad V^i \rightarrow V^i + b^i \, , \qquad \varphi \rightarrow \varphi + 2 b^k X_k + b^k k_{kl} b^l \,, \label{shiftsym}
\ee
mirroring the shift symmetry \eqref{dftshift} appearing in the $\mathrm{O}(D,D)$ non-Riemannian parametrisations of \cite{Morand:2017fnv}. Note that for $b^i$ proportional to $u^i$, $X_i$ is invariant while $V^i$ and $\varphi$ transform. One can eliminate $\varphi$ using the latter transformation, with $b^i = - \frac{1}{2} u^i \varphi$, or eliminate $V^i$ using $b^i = -V^i$.
Furthermore, one can insert \eqref{ansatz} into the $\Gfour$ ExFT  action \eqref{actionSL5} evaluated on the M-theory solution of the section condition to investigate see how $\varphi$ and $V^i$ appear. We focus on the terms involving derivatives of generalised metric. In the ``potential'' $V$, we have:
\be
\begin{split} 
\frac{1}{8}m^{a c} m^{b d} & \partial_{a b} m_{ e f} \partial_{c d}m^{e f}+\frac{1}{2}m^{a c} m^{b d} \partial_{a b}m^{e f} \partial_{e c} m_{ d f}+\frac{1}{2}\partial_{a b}m^{a c} \partial_{c d}m^{b d} \\&
= -\frac{1}{4} u^i u^j \partial_i k_{kl} \partial_j h^{kl} + \frac{1}{2} u^i u^k \partial_i h^{jl} \partial_j k_{kl} 
- \frac{1}{2} \partial_i u^i \partial_j u^j - \frac{1}{2} \partial_i u^j \partial_j u^i 
\\ & \qquad
+ \frac{1}{2} u^i u^j \partial_i u^k \partial_k X_j
+ \frac{1}{2} h^{ij} u^k \partial_i u^l ( \partial_j k_{kl} - \partial_k k_{jl} )
\end{split} 
\ee 
while the ``kinetic term'' involves:\footnote{Here $D_\mu = \partial_\mu - \mathcal{L}_{\Aa_\mu}$ with the generalised Lie derivative acting as explained in the text above.}
\be
\frac{1}{4} 
D_\mu m^{ab} D^\mu m_{ab} = \frac{1}{4} D_\mu k_{ij} D^{\mu} h^{ij} + \frac{1}{2} D_\mu u^i D^\mu X_i
- k_{ij} D_\mu V^i D^\mu u^j + \frac{1}{2} \varphi k_{ij} D_\mu u^i D^\mu u^j \,.
\ee
Before discussing the interpretation of this, note that for these non-Riemannian backgrounds the time-like direction is in what is normally called the ``internal space'' and the external space is Euclidean (this must be the case as in the examples we looked at, we obtained the non-Riemannian background by U-dualising from a generalised metric in $\Gfour/\mathrm{SO}(2,3)$, i.e. including the timelike direction in the extended space). Thus the names ``potential'' and ``kinetic term'' are misnomers inherited from the more usual situation where time is in the external space. The fact that $V^i$ and $\varphi$ drop out of the action involving internal derivatives implies that there are no time derivatives for these fields and so their canonical momentum will vanish. This leads to (in the language of canonical quantisation) a first class constraint which we normally associate with a gauge symmetry. The local symmetry is the shift symmetry we have identified and thus leads to the conclusion that there are fewer physical degrees of freedom than expected relative to the usual case. This should not be a surprise since having non-Riemannian directions will mean that forms will also have fewer degrees of freedom. 

We have seen previously that, in the maximally non-Riemannian case, all the propagating degrees of freedom were projected out and that one should think of the coset as $G/G$. Now that we have some subset of non-Riemannian directions one will naturally have fewer degrees of freedom, as indicated by the presence of the shift symmetry and the first class constraint on $V^i$. Constructing the coset description for each case with a different non-Riemannian structure is an invidious task that we leave for future work. The general story is that $H$ enhances as more dimensions become non-Riemannian. One should also note that from the ``kinetic terms'' ie. the action containing derivatives of the external space on $m_{ab}$ there are additional constraints from the equations of motion, these will be second class. A full detailed analysis of the constraint structure is beyond this paper.  What we wish to emphasise is that there is a reduction in the form degrees of freedom due to the non-Riemannian nature of the space which may be seen from the absence of time derivative terms for $V^i$. The other (second class) constraints maybe viewed as providing a restriction on how a non-Riemannian space may be fibred  over some other space. As such this is model provides a fascinating playground for studying different aspects of how non-Riemannian spaces are embedded in ExFT. 

In spite of the above discussion, we can continue to make use of a $\Gfour / \mathrm{SO}(2,3)$ description for the generalised vielbein. This is perfectly fine so long as we keep in mind the additional shift symmetry (\ref{shiftsym}) that will ultimately lower the degrees of freedom in the coset by enhancing $H$.
Then, to construct the generalised vielbein, let $\bar a, \bar b$ be flat five-dimensional indices, and $\bar \imath, \bar \jmath$ be three-dimensional flat indices. 
The flat generalised metric can be taken to be 
\be
\eta_{ \bar a \bar b} = \begin{pmatrix} \eta_{\bar \imath \bar \jmath} & 0 & 0 \\ 0 & \sigma & 0 \\ 0 & 0 & - \sigma \end{pmatrix} \,,
\ee
where $\sigma = \pm 1$ and $\eta_{\bar \imath \bar \jmath}$ is the three-dimensional Minkowskian metric. A family of generalised vielbein $E^{\bar a}{}_a$, satisfying $m_{ab} = E^{\bar a}{}_a E^{\bar b}{}_b \eta_{\bar a \bar b}$, is provided by 
\be
E^{\bar a}{}_a = \begin{pmatrix} k^{\bar \imath}{}_i & -k^{\bar \imath}{}_j V^j \\ \alpha X_i & \frac{1}{2} \left( \alpha \varphi + \frac{\sigma}{\alpha} \right) - \alpha X^k V_k \\ \alpha X_i & \frac{1}{2} \left( \alpha \varphi - \frac{\sigma}{\alpha} \right) - \alpha X^k V_k \end{pmatrix} \,.
\ee
Here, $\alpha$ is an arbitrary non-zero constant and $k^{\bar\imath}{}_i k^{\bar \jmath}{}_j \eta_{\bar \imath \bar \jmath} = k_{ij}$ with $k^{\bar \imath}{}_i u^i = 0$.

We can also write down the big generalised metric, which will have the parametrisation
\be
M_{ab,cd} 
= \pm (U_V)^T 
\begin{pmatrix}
\varphi k_{ik} - X_i X_k & k_{ki} X_l - k_{li} X_k\\
k_{ik} X_j - k_{jk} X_i & k_{ik} k_{jl} - k_{il} k_{jk} 
\end{pmatrix}
U_V 
\,,\quad U_V^{ab}{}_{cd} = \begin{pmatrix}
\delta^i_k &  0 \\
-\delta^{[i}_k V^{j]} & \delta^{[ij]}_{kl}
\end{pmatrix} \, ,
\ee
where the sign factor will depend on the signature of $k_{ij}$ and on $\sigma$.

Now, consider the example non-Riemannian generalised metric \eqref{exGOm} we generated above by taking the Gomis-Ooguri limit (or equivalently those appearing after U-dualising the M2 supergravity solution, \eqref{exUm}).
This embeds into the general form of the little metric \eqref{mgeneral} as a non-Riemannian parametrisation with:
\be
k_{ij} = \begin{pmatrix} \eta_{\alpha \beta} & 0 \\ 0 & 0 \end{pmatrix} \,,\quad
\chi_i  = \begin{pmatrix} 0 &0 &0 &  1 \end{pmatrix} \,,\quad
\tilde\varphi = -2\mu \,.
\label{mex}
\ee
Alternatively, one could fix $X_i = \chi_i$, $\tilde \varphi = 0$ and write it in the form \eqref{ansatz} in terms of the following dualised three-form
\be
V^i = \begin{pmatrix} 0 & 0 & 0 & \mu \end{pmatrix} \,,
\ee
which implies that we have $C_{t12} = -\mu$, i.e. this three-form is only defined with legs in the three-dimensional space orthogonal to the zero vector of $k_{ij}$. 
This form of the solution is closest to the DFT non-Riemannian parametrisation \eqref{GOH} with a non-vanishing $B$-field. 

\subsubsection*{A quick glance at IIB parametrisations} 

We could approach the issue similarly in a IIB solution of the section condition, writing
\be
m_{ab} = \begin{pmatrix} 
k_{ij} & \chi_{i\beta} \\
\chi_{j\alpha} & H_{\alpha \beta} 
\end{pmatrix} \,,
\ee
with $i,j=1,2,3$ the spacetime indices and $\alpha,\beta=1,2$ the $\mathrm{SL}(2)$ indices.
We have the unit determinant constraint
\be
\mathrm{det}\,k \,\mathrm{det}\,H + \varepsilon^{\alpha \beta} \varepsilon^{\gamma \delta} \eta^{ijk} \eta^{lmn} \left( - \frac{1}{2} H_{\alpha \gamma} \chi_{i \beta} \chi_{l \delta} k_{jm} k_{kn}
+ \frac{1}{4} k_{il} \chi_{j \gamma} \chi_{k \delta} \chi_{m \alpha} \chi_{n \beta} \right) = 1 \,.
\ee
We shall not study this in detail here. It is clear that one can take $k_{ij}$ to be non-invertible, meaning there is no standard spacetime metric. Examples could be found by taking Gomis-Ooguri-type limits of the F1 and D1 SUGRA solutions.
However, there are also ``non-Riemannian'' possibilities that do not involve assuming lack of invertibility of $k_{ij}$ and $H_{\alpha \beta}$. 
Indeed, the example \eqref{mex} (actually, one could in effect interpret this as a Gomis-Ooguri limit for the D$(-1)$ SUGRA solution) corresponds here to 
\be
k_{ij} = \begin{pmatrix} 
- 1 & 0 & 0 \\ 0 & 1 & 0 \\ 0 & 0 & 1 
\end{pmatrix} \,,\quad
\chi_{i\alpha} = 0 \,,\quad
H_{\alpha \beta} = \begin{pmatrix}
0 &  1 \\
1 & -2\mu
\end{pmatrix}\,, 
\ee 
in which the three-by-three and two-by-two blocks are invertible, but, as we explained before, the factor $H_{\alpha \beta}$ does not admit the standard parametrisation in terms of the dilaton and RR zero form as in \eqref{mIIBparam}. Thus not only the geometry, as encoded in the relationship between $k_{ij}$ and the spacetime metric, but the information about the string coupling can be modified in a generic parametrisation of the generalised metric.

\subsection{Reduction to $\mathrm{O}(3,3)$}
\label{redtoO33} 

We now reduce the $\Gfour$ non-Riemannian parametrisation \eqref{ansatz} to $\mathrm{O}(3,3)$ to show that it becomes a $(1,1)$ non-Riemannian parametrisation in DFT. 
The reduction from the $\Gfour$ ExFT to the $\mathrm{O}(3,3)$ DFT follows \cite{Berman:2011cg,Thompson:2011uw} (see also the appendix of \cite{Blair:2018lbh}).
The $\mathbf{10}$ of $\Gfour$ reduces to the $\mathbf{6} \oplus \mathbf{4}$ of $\mathrm{O}(3,3)$, while the $\mathbf{5}$ of $\Gfour$ becomes that $\mathbf{4} \oplus \mathbf{1}$.
The $\mathbf{4}$ is a Majorana-Weyl spinor representation. 
The generalised metric $\gM_{ab,cd}$ then leads to the $\mathrm{O}(3,3)$ generalised metric, $\cH_{MN}$, a Majorana-Weyl spinor of RR fields, $C_I$, where $I$ is a four-dimensional spinor index, and the generalised dilaton, $\gendil$. 
It is convenient to phrase the reduction in terms of the little metric: this gives rise to the $\mathrm{O}(3,3)$ generalised metric in the form of a $4\times 4$ symmetric matrix $h_{IJ}$ carrying spinorial indices. Explicitly, we have
\be
m_{ab} = 
\begin{pmatrix} 
e^{-2\gendil/5} h_{IJ} + \eta e^{8\gendil/5} C_I C_J & \eta e^{8\gendil/5} C_I \\
\eta e^{8\gendil/5} C_J & \eta e^{8\gendil/5} 
\end{pmatrix} 
\label{mtoh}
\ee
We include $\eta = \pm 1$ to allow the description of ``timelike'' reductions (see appendix \ref{appred} for the details of how this features in reductions of the usual Riemannian parametrisations). It is convenient to think of the five dimensional index $a$ as being split $a=(I,4)$. 

Recall that the parametrisation of the large generalised metric is given by $\gM_{ab,cd} = \pm ( m_{ac} m_{bd} - m_{ad} m_{bc} )$, where the sign is chosen such that the generalised line element $\gM_{MN} dY^M dY^N \sim + \phi_{ i j} dY^i dY^j + \dots$ on choosing a section condition solution $\partial_i \neq 0$. 
In order to pick out the components of $\gM_{MN}$ corresponding to the $S\mathrm{O}(3,3)$ generalised metric, it is convenient to work with this generalised line element, for which we have
\be
\begin{split}
\gM_{MN} dY^M dY^N & = \pm \bigg( \left( \frac{1}{2} e^{-4d/5} h_{IK} h_{JL} + \eta e^{6d/5} C_I C_K h_{JL} \right) dY^{IJ} dY^{KL} 
\\ & \qquad\qquad+ 2 \eta e^{6d/5} h_{IK} C_J dY^{IJ} dY^{K4} + \eta e^{6d/5} h_{IJ} dY^{I4} dY^{J4} 
\bigg)\,,
\end{split}
\ee
where the 10 $\Gfour$ coordinates $Y^M$ split into four spinorial coordinates $Y^{I4}$ and six coordinates $Y^{IJ}$ carrying an antisymmetric pair of spinor indices. These are related to the usual doubled coordinates by $Y^M = \frac{1}{2\sqrt{2}} \gamma^M{}_{IJ} Y^{IJ}$, where $\gamma^M$ and $\gamma_M$ denote the off-diagonal blocks of the full $\mathrm{O}(3,3)$ gamma matrices \cite{Berman:2011cg}. Writing $I=(i, \sharp)$ (where really $\sharp$ can be identified with $a=5$ index of the original $\Gfour$ ExFT) means that $Y^M = ( Y^i, \tilde Y_i ) = ( Y^{i\sharp} , \frac{1}{2} \eta_{ijk} Y^{jk} )$, such that the components of the usual DFT generalised metric are given by
\be
\begin{split}
\cH_{ ij }  = \pm (h_{ij} h_{\sharp \sharp} -  h_{ i\sharp} h_{ j \sharp } ) \,,\quad
\cH_{i}{}^j  = \pm \eta^{jkl} h_{ik} h_{l\sharp} \,,\quad
\cH^{ij}  = \pm \frac{1}{2} \eta^{ikl} \eta^{jmn} h_{km} h_{ln} \,.
\end{split}
\label{Hfromh}
\ee
Let us carry this process out for the non-Riemannian parametrisation in the original form \eqref{mgeneral}, that is, let us write
\be
m_{ab} = \begin{pmatrix}
 k_{\hat \imath \hat \jmath} & \chi_{\hat \imath} \\
 \chi_{\hat \jmath} & \tilde\varphi
\end{pmatrix} 
\ee
where now $a=(\hat \imath, 5)$ in order to rewrite the four-dimensional indices as $\hat \imath = (i,4)$.
We assume that $k_{44} \neq 0$ and, for simplicity, that $k_{44} > 0$ so the reduction is (in some sense) spacelike. We can then write a Kaluza-Klein style decomposition for the degenerate matrix $k_{\hat \imath \hat \jmath}$,
\be
k_{ \hat \imath \hat \jmath} = \begin{pmatrix}
 \tilde k_{ij} + \frac{1}{k_{44}} k_{i4} k_{j4} & k_{i 4} \\ 
  k_{j 4} & k_{44} 
\end{pmatrix} \,.\ee
Then $\det k = k_{44} \det \tilde k$ and so $\det \tilde k = 0$. The unit determinant condition \eqref{mdetconstraint} on the generalised metric implies 
\be
-\frac{1}{2} \eta^{ijk} \eta^{lmn} \tilde \chi_i  \tilde \chi_l k_{44} \tilde k_{jm} \tilde k_{kn} = 1\,,
\ee
where $\tilde \chi_i = \chi_i - k_{i 4} \chi_4 / k_{44}$. The null vector becomes
\be
u^i = - \frac{1}{2} \eta^{imn} \eta^{jkl} \tilde \chi_j k_{44} \tilde k_{km} \tilde k_{lm} \,,
\quad 
u^4 = - k_{i 4} u^i / k_{4 4}\,,
\ee
with $\tilde k_{ij} u^i = 0$, $\tilde \chi_i u^i = 1$. Using \eqref{mtoh} we find explicit expressions for the generalised dilaton, the RR fields, and the spin generalised metric:
\be
e^{8d/5} = k_{44}\,,\quad
C_I = \begin{pmatrix}k_{44}^{-1} k_{i4} \\ k_{44}^{-1} \chi_4 \end{pmatrix} 
\,,\quad
h_{IJ} = ( k_{44} )^{1/4} \begin{pmatrix}
\tilde k_{ij} & \tilde \chi_i \\ 
\tilde \chi_j & \tilde\gamma 
\end{pmatrix} \,,
\ee
where $\tilde \gamma \equiv \tilde\varphi - k_{44}^{-1} ( \chi_4 )^2$.
This leads, via \eqref{Hfromh}, to the components of the vector generalised metric:
\be
\begin{split}
\mathcal{H}_{ij}  = \pm ( k_{44} )^{1/2} \left( \tilde\gamma \tilde k_{ij} - \tilde \chi_i \tilde \chi_j \right)\,,\quad
\mathcal{H}_i{}^j  =\pm ( k_{44} )^{1/2} \eta^{jkl} \tilde k_{ik} \tilde \chi_l  \,,\quad
\mathcal{H}^{ij}  = \pm ( k_{44} )^{1/2} \frac{1}{2} \eta^{ikl} \eta^{jpq} \tilde k_{kp} \tilde k_{lq} \,,
\end{split} 
\label{tosolve}
\ee
where $\pm$ corresponds to the choice of sign in relating the parametrisation of $\gM_{ab,cd}$ to that of the little metric.

We note that 
\be
\mathcal{H}^{ij} \tilde k_{jk} Y^k = 0 
\ee
for arbitrary vector $Y^k$. Now, $\tilde k_{ij}$ has only one zero vector $u^i$, so its kernel has dimension one, and thus its image has dimension two.
This implies that $\mathcal{H}^{ij}$ has a two-dimensional kernel.
We therefore identify it directly with the degenerate matrix $H^{ij}$ of the DFT parametrisation.
We also see that $\mathcal{H}_i{}^i = 0$, so the trace vanishes, $\mathcal{H}_M{}^M =0$.
Thus the generalised metric obtained in the reduction is necessarily of type $(1,1)$ and not $(2,0)$ or $(0,2)$.

The identification of the other elements of the DFT parametrisation is ambiguous owing to the presence of shift symmetries in the non-Riemannian parametrisation.
One possible choice would be identify $K_{ij}$ with the term in $\mathcal{H}_{ij}$ proportional to $\tilde\chi_i \tilde\chi_j$.
After some work, detailed in appendix \ref{reddetails}, it can be shown that this corresponds to:
\be
\begin{split} 
	H^{ij}  = \pm \frac{1}{2} (k_{44})^{1/2} \eta^{ikl} \eta^{jmn} \tilde k_{km} \tilde k_{ln} \,, \quad
	K_{ij}  = \mp ( k_{44} )^{1/2} \tilde X_i \tilde X_j \,,\\
\end{split}
\label{mtoDFTres1}
\ee
\be
\begin{split} 
	B_{ij}  = \pm \frac{1}{2} (k_{44})^{1/2} \tilde\gamma \eta_{ijk} H^{kl} \tilde X_l \,,\quad
	Z_i{}^j = \pm ( k_{44} )^{1/2} \eta^{jkl} \tilde k_{ik} \tilde X_l \,.
\end{split}
\label{mtoDFTres2}
\ee
In this case, the zero vectors of $K_{ij}$ are those $Y^i$ such that $\tilde X_i Y^i = 0$, while those of $H^{ij}$ are those $X_i$ such that $X_i = k_{ij} Y^j$. This parametrisation amounts to a special choice of $B$-field such that $B_{ij} H^{jk} = 0$.

Regardless of the ambiguity in directly identifying the blocks as in \eqref{mtoDFTres1} and \eqref{mtoDFTres2}, the expressions \eqref{tosolve} allow us obtain the full DFT generalised metric without ambiguities.

Let us check how this works out for the non-Riemannian little metrics \eqref{mex} which corresponded to the Gomis-Ooguri limit or the timelike U-duality of the M2 solution. 
Reducing as above we find
\be
h_{IJ} = \begin{pmatrix} -1 & 0 & 0 & 0 \\ 0 & 1 & 0 & 0 \\ 0 & 0 & 0 & \pm 1 \\ 0 & 0& \pm 1 & -f \end{pmatrix} \,,\quad e^{-2d} = 1\,,
\ee
with vanishing $C_I$. This then leads to (the generalised metric is still Lorentzian so we choose the minus sign in $\pm$ of \eqref{mtoDFTres1})
\be
\cH_{ij} 
= \begin{pmatrix} f & 0 & 0 \\ 0 & - f & 0 \\ 0 & 0 & 1 \end{pmatrix} \,,\quad
\cH^{ij} =
\begin{pmatrix} 0 & 0 & 0 \\ 0 & 0 & 0 \\ 0 & 0 & 1 \end{pmatrix} \,,\quad
\cH_i{}^j 
 = \begin{pmatrix} 0 & \mp 1 & 0 \\ \mp 1 & 0 & 0 \\ 0 & 0 & 0 \end{pmatrix} \,.
\ee 
For the Gomis-Ooguri case, when $f=-2\mu$ and the off-diagonal entries are $+1$, we get exactly the generalised metric obtained in the original Gomis-Ooguri limit, \eqref{GOH}.
For the background resulting from U-duality of the M2, with $f=H$ and $-1$ in the off-diagonal entries, we find that this generalised metric is $-P_\epsilon \tilde{\cH} 
P_\epsilon$, where $\tilde{\cH}_{MN} $ is the generalised metric we obtained in the DFT case by T-dualising on the worldsheet directions of the F1 string, given in \eqref{Hnrex}. 
The geometric $O(3,3)$ transformation
\be
P_\epsilon =
\begin{pmatrix}
A & 0 \\ 0 & A^{-T} 
\end{pmatrix}\,,\quad
A\equiv \begin{pmatrix} 
 0 & 1 & 0\\
-1 & 0 & 0 \\
 0 & 0 & 1\\
\end{pmatrix} \,,
\quad P_\epsilon^T = P_\epsilon^{-1} = - P_{\epsilon}\,,
\ee
corresponds to an interchange of $t$ and $z^1$. Its appearance is to be expected if we consider the U-duality transformation we used on the M2 solution. 
For $\alpha=(t,1,2)$ this acted on the coordinates as
\be
Y^{\prime\alpha 5}  = - Y^{\alpha w} \,,\quad
Y^{\prime\alpha j}  = Y^{\alpha \beta} \,,\quad
Y^{\prime\alpha w}  = Y^{\alpha 5} \,,\quad
Y^{\prime w5}  = Y^{w5} \, 
\ee
so that if we reduce from $\mathrm{SL}(5)$ to $\mathrm{SO}(3,3)$ on the $i=2$ direction, we find this acts on the doubled coordinates as $Y^M = (Y^t,Y^1, \tilde Y_t , \tilde Y_1)$ as $Y^{\prime M} = (P_\epsilon \mathcal{T})^M{}_N$ where $\mathcal{T}^M{}_N$ is the Buscher transformation on the $t,z^1$ directions. Note that $P_\epsilon \mathcal{T} = \mathcal{T} P_\epsilon$.
This result is therefore to be expected. Diagrammatically, we have:
\begin{center}
\begin{tikzcd}
\text{ExFT:} & m_{ab} \arrow[r, "U"] \arrow[d, "\text{reduce}"]
& \tilde m_{ab} \arrow[d, "\text{reduce}"] \\
\text{DFT:} & \cH_{MN} \arrow[r,, "P_\epsilon \mathcal{T}"]
& \tilde{\cH}_{MN} \,.
\end{tikzcd}
\end{center}

\subsection{Embedding Newton-Cartan}
\label{embeddingNC}

We will now offer three variations on a theme of Newton-Cartan.
Recall from section \ref{DFTNewtonCartan} that the Newton-Cartan geometry is described by degenerate matrices $h_{\mu\nu}$, $h^{\mu\nu}$ and their zero vectors $v^\mu$ and $\tau_\mu$. 
Here the index $\mu$ is $d$ dimensional and $h_{\mu\nu}$ and $h^{\mu\nu}$ have rank $d-1$.
We will describe below methods to embed a Newton-Cartan geometry in an $\Gfour$ description for $d=4,3$ and $2$. We will only discuss the generalised metric, i.e the ``internal'' sector of the ExFT. Including the external metric and other degrees of freedom would extend this to putative $11$, $10$ and $9$-dimensional non-relativistic geometries.
The latter case corresponds to the embedding of the Newton-Cartan geometry discussed in the context of string theory in section \ref{DFTNewtonCartan}, so that our description here might be thought of as an M-theory uplift of this. The 10-dimensional case would be obtained in M-theory by U-dualising an 11-dimensional metric along a null isometry direction.
The 11-dimensional case would perhaps be best thought of as needing ExFT for a higher dimensional interpretation, in place of a 12-dimensional Lorentzian geometry.
We leave detailed inquiry into these possibilities, and their role in an potential non-relativistic duality web, for future work.

\subsubsection*{$d=4$ Newton-Cartan: direct non-Riemannian and Riemannian embeddings}

The structure of our non-Riemannian parametrisation, in terms of degenerate metric-like quantities with zero vectors is clearly very similar to that of the Newton-Cartan geometry. 
Indeed, given the Newton-Cartan structure $h_{\mu\nu}, h^{\mu\nu}$, $v^\mu$ and $\tau_\mu$ (we will consider the gauge field $m_\mu$ subsequently), with $\mu,\nu=1,\dots,4$, let us define first
\be
g \equiv - \frac{1}{6} \tau_{\mu_1} \tau_{\nu_2} \eta^{\mu_1 \dots \mu_4} \eta^{\nu_1 \dots \nu_4} h_{\mu_2 \nu_2} h_{\mu_3 \nu_3} h_{\mu_4 \nu_4} \,,
\ee
such that $|g|^{1/2}$ may serve as a density factor. (If we were to embed the Newton-Cartan geometry in a Lorentzian metric with a null isometry as in section \ref{DFTNewtonCartan}, then $g$ is indeed the determinant of this metric.) Then the following non-Riemannian $\Gfour$ parametrisation describes this geometry: 
\be
m_{ab} = \frac{g}{|g|} \begin{pmatrix}
|g|^{-2/5} h_{\mu\nu} & |g|^{1/10} \tau_\mu \\
|g|^{1/10} \tau_\nu & 0 
\end{pmatrix} \,,\quad
m^{ab} = \frac{g}{|g|}\begin{pmatrix}
|g|^{2/5}  h^{\mu\nu} & - |g|^{-1/10} v^\mu \\
- |g|^{-1/10} v^\nu & 0 
\end{pmatrix} 
\ee
where we have to include a sign factor $\frac{g}{|g|}$, in order that $\det m_{ab} = 1$. 
(Note the form $m_{ab}$ itself resembles that of a five-dimensional metric with a null isometry.)

Curiously, we can include the Newton-Cartan gauge field $m_\mu$ in the form of conjugation by the following $\Gfour$ element (which seems natural from the point of view of the index structure of $m_\mu$):
\be
U_W^a{}_b = \begin{pmatrix} \delta^\mu{}_\nu & 0 \\ W_\nu & 1 \end{pmatrix}\,,
\quad
W_\mu \equiv - |g|^{-1/2}  m_\mu \,,
\ee
such that
\be
\begin{split}
\tilde m_{ab} & = (U_W)^c{}_a m_{cd} (U_W)^d{}_b = \frac{g}{|g|}\begin{pmatrix}
|g|^{-2/5}\bar h_{\mu\nu} & |g|^{1/10} \tau_\mu \\
|g|^{1/10} \tau_\nu & 0 
\end{pmatrix} \,,\\
\tilde m^{ab}& = (U^{-1}_W)^a{}_c (U^{-1}_W)^b{}_d m^{cd} = \frac{g}{|g|}\begin{pmatrix}
|g|^{2/5} h^{\mu\nu} & - |g|^{-1/10} \hat v^\mu \\
- |g|^{-1/5} \hat v^\nu & 2 |g|^{3/5} \tilde \Phi
\end{pmatrix} 
\end{split}
\ee
Ordinarily, the factorisation involving $U_W$ would be associated with a trivector $\Omega^{\mu\nu\rho}$ defined by $W_\mu \equiv \frac{1}{3!} \eta_{\mu\nu\rho\sigma} \Omega^{\nu\rho\sigma}$.
The presence of the trivector, analogous to the bivector which appears in $\mathrm{O}(d,d)$, is usually associated to non-geometry (for instance, it provides a potential for non-geometric fluxes \cite{Blair:2014zba}).
This may signal here that we should really interpret this generalised metric in terms of a ``dual'' solution of the section condition. Furthermore, $\tilde m_{ab}$ now actually represents a Riemannian parametrisation as $\det \bar h = - 2 \Phi e^2 \neq 0$. 
We leave a detailed understanding of this geometry, and its local symmetries, for future work.

\subsubsection*{$d=3$ Newton-Cartan: U-dualising the Lorentzian metric} 

The second approach mimics what we did in section \ref{DFTNewtonCartan}. We start with the form of the Lorentzian metric \eqref{metric1} with null isometry, embed this into the $\Gfour$ ExFT and then U-dualise. That means we are thinking of \eqref{metric1} as describing an 11-dimensional metric, which may provide a route to a 10-dimensional Newton-Cartan geometry. 
We focus on a four-dimensional part of the metric: 
\be
\phi_{ij} = \begin{pmatrix} \bar h_{\mu\nu} & \tau_\mu \\ \tau_\nu & 0 \end{pmatrix} \,,
\ee
where now $\mu$ is a three-dimensional index, and $\bar h_{\mu\nu} =h_{\mu\nu} - 2 \tau_{(\mu} m_{\nu)}$ as before. 
We insert this into the Lorentzian parametrisation of the little metric \eqref{mparam} with $\lambda = 1$ and $(-1)^t = - 1$.
Then we U-dualise via \eqref{UBuscher} with $n_i = ( n_\mu, 0 )$. This corresponds to a U-dualisation on the null isometry direction $u$ and two of the three-directions indexed by $\mu$. (Strictly speaking, we should impose that two of these directions are also isometries.) 
The resulting generalised metric is
\be
m_{ab} = |\phi|^{-2/5}
\begin{pmatrix}
 \bh_{\mu\nu} + n_\mu n_\nu ( n^\rho n^\sigma \bh_{\rho\sigma} - |\phi| ) - 2 n^\rho \bh_{\rho ( \mu } n_{\nu ) }  &   \tau_\mu - \tau^\rho n_\rho n_\mu  &  \bh_{\mu\rho}n^\rho - n_\mu n^\rho n^\sigma \bh_{\rho \sigma} \\ 
  \tau_\mu - \tau^\rho n_\rho n_\mu  &  0  &  n^\rho \tau_\rho \\
   \bh_{\nu\rho}n^\rho - n_\nu n^\rho n^\sigma \bh_{\rho \sigma}  &  n^\rho \tau_\rho &  \bh_{\mu\nu} n^\mu n^\nu
\end{pmatrix} \,.
\label{NClittlem1}
\ee
This becomes non-Riemannian for $n_\mu = \tau_\mu$ and $n^\mu = -v^\mu$ or $n^\mu = - \hv^{\mu}$. 
For instance, in the latter case we have
\be
k_{ij} = |\phi|^{-2/5} \begin{pmatrix} \bh_{\mu\nu} + ( 2 \tilde \Phi - |\phi| ) \tau_\mu \tau_\nu & 0 \\ 0 & 0 \end{pmatrix} 
\,,
\quad 
\chi_i = |\phi|^{-2/5} \begin{pmatrix} 0 \\ 1 \end{pmatrix}
\,,\quad
\tilde\varphi = - |\phi|^{-2/5} 2 \tilde \Phi \,.
\ee

\subsubsection*{$d=2$ Newton-Cartan: uplift from $\mathrm{O}(3,3)$ DFT} 

The final approach we take is to start with the $\mathrm{O}(3,3)$ description of the Newton-Cartan geometry, which has a concrete origin in string theory via the dualisation procedure, and uplift this. So, we start again with the Newton-Cartan variables $h_{\mu\nu}$, $h^{\mu\nu}$, $\tau_\mu$, and $v^\mu$, where now $\mu$ is a two-dimensional index. 
The Newton-Cartan geometry is especially simple as $h_{\mu\nu}$ and $h^{\mu\nu}$ have rank 1, and can be written in general as
\be
h_{\mu\nu}  = e^2 h_\mu h_\nu\,,\quad h^{\mu\nu} = \frac{1}{e^2} h^\mu h^\nu\,,
\ee
where
\be
h_\mu \equiv \varepsilon_{\mu\nu} v^\nu \,,\quad
h^\mu \equiv \varepsilon^{\mu\nu} \tau_\nu \,,
\ee
such that $h_\mu v^\mu = 0 = h^\mu \tau_\mu$, $h^\mu h_\mu = -1$ and the completeness relation holds, $h^{\mu \rho} h_{\rho \nu} - v^\mu \tau_\nu = \delta^\mu_\nu$. Here our conventions are that $\varepsilon_{12} = 1$, $\varepsilon^{12} = 1$ and so $\varepsilon_{\mu \rho} \varepsilon^{\nu \rho}  = + \delta_\mu^\nu$
($\varepsilon^{\mu\nu}$ rather than $\eta^{\mu\nu}$ to denote the alternating symbol as the latter may be confused for a metric).
We have chosen to parametrise $h_{\mu\nu}$ in terms of a positive function $e^2 > 0$, as required by going back to the Lorentzian metric from which the Newton-Cartan geometry can be obtained by null duality, which has $\det g = - e^2$, using the formula \eqref{detgNC}.

We can then reverse engineer the $\Gfour$ generalised metric which reduces to the $\mathrm{O}(3,3)$ Newton-Cartan generalised metric \eqref{NCGM}.
Using \eqref{tosolve}, we find that the $\mathrm{SL}(5)$ ExFT uplift of the Newton-Cartan geometry is described by the following non-Riemannian parametrisation of the little metric:
\be
m_{ab} = \begin{pmatrix} 
- e^{-4/5} \tau_\mu \tau_\nu & 0 & 0 &  e^{6/5} \varepsilon_{\mu\rho} \hat v^\rho \\
0 & e^{-4/5} & 0 & 0 \\
0 & 0 & e^{-4/5} & 0 \\
e^{6/5} \varepsilon_{\nu\rho} \hat v^\rho & 0 & 0 & - e^{6/5} 2 \tilde \Phi  
\end{pmatrix} \,.
\label{NClittlem}
\ee
Note that $\tau_\mu \tau_\nu = e^2\varepsilon_{\mu\rho} \varepsilon_{\nu \sigma} h^{\rho \sigma}$.
It is possible to find U-duality transformations that take it to a Riemannian background - we describe a couple of possibilities in appendix \ref{app:NCU}.

\section{Discussion}

Exceptional Field Theory was constructed to reproduce ordinary supergravity but with the additional feature that it could manifestly include the different perspectives of U-duality related geometries in M-theory. Since its inception, some surprises have emerged, and the full spectrum of theories and backgrounds which can be accommodated in the ExFT (and DFT) framework is still being uncovered. 
ExFT (and DFT) can describe many variants of supergravity theories.
By focusing on non-Riemannian parametrisations, we extend the range of DFT/ExFT further, finding that it can accommodate non-relativistic theories and theories seemingly without a standard dynamical gravity at all in the form of the maximally non-Riemannian solutions which gave either chiral string theory (in DFT) in a beta-gamma sense \cite{Nekrasov:2005wg} or a topological three-dimensional theory (in $\Geight$ ExFT).

In this paper we have extended the results of \cite{Lee:2013hma,Ko:2015rha,Morand:2017fnv,Cho:2018alk} and initiated the study of non-Riemannian backgrounds appearing in exceptional field theory. We have shown how the $\Geight$ ExFT contains a maximally non-Riemannian solution with no moduli. In this background the ExFT becomes the topological theory previously described by Hohm and Samtleben \cite{Hohm:2018ybo} as a truncation of ExFT. For the $\Gfour$ ExFT, we have shown how to parametrise the generalised metric in order to obtain theories of non-relativistic type, related to the Gomis-Ooguri scaling limit and to Newton-Cartan type geometries.

The next stage is to understand these maximally non-Riemannian backgrounds and more of their properties. One can perhaps think of there being a full ``package'' associated to DFT/ExFT which is by now well understood and incorporates for instance geometric notions (such as connections and curvatures), supersymmetry, Scherk-Schwarz compactifications, and aspects of the descriptions of strings and branes \cite{Berman:2014jsa,Berman:2014hna}. This package should be brought to bear on understanding theories of non-Riemannian geometry and their role in string and M-theory.

A natural step is to examine sigma models in these backgrounds and determine their quantum consistency.
An immediate observation is that the doubled sigma model in such a background contains chiral theories \cite{Morand:2017fnv}, and thus the quantum consistency of the partition function is highly constrained \cite{Nekrasov:2005wg}.  
The situation is similar for the exceptional sigma model of \cite{Arvanitakis:2017hwb,Arvanitakis:2018hfn}, which provides a way to describe strings and D1 branes in ExFT non-Riemannian backgrounds (likely reproducing and generalising the results of the very recent paper \cite{Kluson:2019ifd}). Furthermore one could study the general brane actions of \cite{Sakatani:2017vbd} in such backgrounds.

\section*{Acknowledgements}

David Berman is supported by STFC grant ST/L000415/1, ``String Theory, Gauge Theory and Duality''. 
CB is supported by an FWO-Vlaanderen Postdoctoral Fellowship, and this work was furthermore supported in part by the Belgian Federal Science Policy Office through the Interuniversity Attraction Pole P7/37, in part by the FWO-Vlaanderen through the projects G020714N and G006119N, and in part by Vrije Universiteit Brussel through the Strategic Research Program ``High-Energy Physics''. 
We would like to thank the Corfu meeting, ``Dualities and Generalized Geometries'' part of the  COST Action MP1405, for providing the initial stimulus for this project.
We would also like to thank Chris Hull, Gianluca Inverso, Emanuel Malek, Diego Marqu{\'e}s, Niels Obers, Jeong-Hyuck Park and Henning Samtleben for useful discussions on different aspects of this work.

\appendix 

\section{Reduction from $\Gfour$ to $\mathrm{O}(3,3)$} 
\label{appred}

Here we summarise the results of reducing the Riemannian parametrisations of the $\Gfour$ generalised metric to $\mathrm{O}(3,3)$. The relationships between the exotic variants of SUGRA are detailed in figure \ref{exoticdiagram}.

\subsection{M-theory/IIA parametrisation}

We start with the M-theory parametrisation of the generalised metric, \eqref{mparam}.
The usual reduction from M-theory to IIA, allowing for the possibility of reducing on a timelike circle, gives the generalised metric \eqref{mtoh} with
\be
h_{IJ} = \begin{pmatrix} 
\lambda |g|^{-1/2} g_{ij} & \lambda |g|^{-1/2} g_{ik} B^k \\
\lambda |g|^{-1/2} g_{jk} B^k & \eta (-1)^t |g|^{1/2} + \lambda |g|^{-1/2} g_{kl} B^k B^l 
\end{pmatrix} \,,\quad
C_I = \begin{pmatrix} C_i \\ \frac{1}{3!} \eta^{ijk} ( C_{ijk} - 3 C_i B_{jk} ) \end{pmatrix} \,,
\ee
and the usual generalised dilaton $e^{-2\gendil} = e^{-2\Phi} \sqrt{|g|}$. Here $B^i \equiv \frac{1}{2} \eta^{ijk} B_{jk}$ is the $B$-field, $C_{ijk}$ is the RR three-form and $C_i$ the RR one-form. The number $t$ denotes the number of timelike directions present in $g_{ij}$, which is the string frame metric. The sign factor $\eta$ takes into account whether the reduction is spacelike ($\eta = +1$) or timelike ($\eta = -1$).
Using the result \eqref{Hfromh}, with the sign factor $\pm$ replaced by $(-1)^t \eta\lambda$, we find that
\be
\mathcal{H}_{MN} = \lambda \eta \begin{pmatrix}
\lambda \eta g_{ij} - B_{ik} g^{kl} B_{lj} & B_{ik} g^{kj} \\
- g^{ik} B_{kj} & g^{ij} 
\end{pmatrix}  \,.
\ee
The following sign factors are possible:
\begin{itemize}
\item $\lambda=+1$, $\eta=+1$ corresponds to a spacelike reduction from conventional M-theory to conventional IIA, giving the usual DFT generalised metric
\item $\lambda=+1$, $\eta=-1$ corresponds to a timelike reduction from conventional M-theory to exotic IIA${}^{-+}$, giving the generalised metric for the exotic DFT${}^-$ constructed in \cite{Blair:2016xnn}. 
\item $\lambda = -1$, $\eta = +1$ corresponds to a spacelike reduction from the exotic M${}^-$ theory to IIA${}^{--}$, giving again the DFT${}^-$ generalised metric.
\item $\lambda - 1$, $\eta=-1$ corresponds to a timelike reduction from M${}^-$ to IIA${}^{+-}$, giving the usual DFT generalised metric. Here though, the RR fields will have the wrong sign kinetic term
\end{itemize}

\subsection{IIB parametrisation} 

We can start with the IIB parametrisation \eqref{mIIBparam} and rewrite in terms of a string frame metric $g_{ij} = e^{\Phi/2}\phi_{ij}$. The sign factor $\eta$ is identified with $\sigma_F$.
It is convenient to also raise all the spinorial indices in the decomposition of the little metric \eqref{mtoh}. This reflects the fact that the RR spinors in a IIB parametrisation of DFT have opposite chirality to those in a IIB parametrisation. Thus, we write
\be
h^{IJ} = \begin{pmatrix} 
\sigma_F \sigma_D (-1)^t |g|^{1/2} g^{ij} + \sigma_D |g|^{-1/2} B^i B^j & \sigma_D |g|^{-1/2} B^i \\
\sigma_D {|g|}^{-1/2} B^j & \sigma_D |g|^{-1/2} 
\end{pmatrix}  \,,\quad
C^I = \begin{pmatrix} \frac{1}{2} \eta^{ijk} ( C_{jk} + C_0 B_{jk} ) \\ C_0 \end{pmatrix} \,,
\ee
and the usual generalised dilaton $e^{-2\gendil} = e^{-2\Phi} \sqrt{|g|}$. Again $B^i \equiv \frac{1}{2} \eta^{ijk} B_{jk}$ is the $B$-field, while $C_{ij}$ is the RR two-form and $C_0$ the RR zero-form. The number $t$ denotes the number of timelike directions present in $g_{ij}$. 
We now get for the generalised metric (the coordinates are now given by $Y^M = \eta^{MN} \frac{1}{2\sqrt{2}} \gamma_N{}^{IJ} Y_{IJ}$), using that the $\pm$ factor is given by $(-1)^t$ in IIB parametrisations of the $\Gfour$ ExFT, 
\be
\mathcal{H}_{MN} = \sigma_F \begin{pmatrix}
\sigma_F g_{ij} - B_{ik} g^{kl} B_{lj} & B_{ik} g^{kj} \\
- g^{ik} B_{kj} & g^{ij} 
\end{pmatrix}  \,.
\ee
We thus see that when $\sigma_F = +1$, we get the usual generalised metric of DFT, while when $\sigma_F=-1$, we get that of DFT${}^-$, as expected to describe IIB${}^{-+}$.

\subsection{DFT non-Riemannian parametrisation from $\Gfour$ non-Riemannian parametrisation}
\label{reddetails}

In this subappendix, we explain how the equations \eqref{tosolve}, namely,
\be
\begin{split}
\mathcal{H}^{ij}  = \pm ( k_{44} )^{1/2} \frac{1}{2} \eta^{ikl} \eta^{jpq} \tilde k_{kp} \tilde k_{lq} \,,\quad
\mathcal{H}_i{}^j  =\pm ( k_{44} )^{1/2} \eta^{jkl} \tilde k_{ik} \tilde \chi_l  \,,\quad
\mathcal{H}_{ij}  = \pm ( k_{44} )^{1/2} \left( (\tilde \gamma \tilde k_{ij} - \tilde \chi_i \tilde \chi_j \right)\,,
\end{split} 
\ee
can be solved to express the blocks of the $O(3,3)$ DFT non-Riemannian parametrisation in terms of the quantities $\tilde k_{ij}$, $\tilde \chi_i$, $\tilde\gamma$, that result from the decomposition of the $\Gfour$ generalised metric.
We choose to take
\be
H^{ij} = \pm \frac{1}{2} (k_{44})^{1/2} \eta^{ikl} \eta^{jmn} \tilde k_{km} \tilde k_{ln} \,,
\label{HfromAbove}
\ee
which is the unique choice for $H^{ij}$, and 
\be
K_{ij}  = \mp ( k_{44} )^{1/2} \tilde \chi_i \tilde \chi_j \,,
\label{KfromAbove}
\ee
which is not unique but is particularly natural from the reduction point of view as it is clearly of rank 1. 
Other choices of $K_{ij}$ can be obtained using shift symmetries. 
Having committed to this choice, we note that $\mathcal{H}_{i}{}^j K_{jk} = 0 = H^{ij} \mathcal{H}_{j}{}^k$.
Now, we let $Z_i{}^j \equiv X_i^a Y^j_a - \bar X_i^{\bar a} \bar Y^j_{\bar a}$ in terms of the preferred basis of zero vectors of $K_{ij}$ and $H^{ij}$.
As $\mathcal{H}_{i}{}^j = Z_i{}^j + B_{ik} H^{kl}$ we see that $B_{ij} H^{jk} K_{kl} = 0$.
Hence
\be
\begin{split} 
B_{ij} & = ( K_{ik} H^{kl} + X_i^a Y_a^l + \bar X_i^{\bar a}  \bar Y^l_{\bar a} )B_{lm} (  K_{jp} H^{pm} + X_j^a Y_a^m + \bar X_j^{\bar a}  \bar Y^m_{\bar a} )\\
&= ( X_i^a Y_a^l + \bar X_i^{\bar a}  \bar Y^l_{\bar a} )B_{lm} (  X_j^a Y_a^m + \bar X_j^{\bar a}  \bar Y^m_{\bar a} )
\end{split}
\ee
and so in fact $H^{ij} B_{jk} = 0$. 
We conclude therefore that
\be
Z_i{}^j = \pm ( k_{44} )^{1/2} \eta^{jkl} \tilde k_{ik} \tilde \chi_l \,.
\label{ZfromAbove}
\ee
We then need to solve 
\be
\pm ( k_{44} )^{1/2}  \tilde \gamma \tilde k_{ij} = B_{ik} Z_j{}^k + B_{jk} Z_i{}^k \,,
\label{hereweare}
\ee
with $Z$ as in \eqref{ZfromAbove}.
Contracting with $u^i$, the left-hand-side vanishes. Writing $B_{ij} = \eta_{ijk} B^k$ we can then show that $\tilde k_{ij} B^j =0$ and hence $B^i = \beta u^i$.
Inserting this back into \eqref{hereweare} it follows that $\beta = -\frac{1}{2} \tilde\gamma$.
Hence, we find
\be
B_{ij} = - \frac{1}{2} \tilde \gamma \eta_{ijk}  u^k 
 = \pm \frac{1}{2} (k_{44})^{1/2} \tilde \gamma \eta_{ijk} H^{kl} \tilde \chi_l \,.
\ee

\section{U-duality of Newton-Cartan uplift}
\label{app:NCU}

In section \ref{embeddingNC} we found an uplift of the $\mathrm{O}(3,3)$ description of Newton-Cartan geometry into the $\mathrm{SL}(5)$ ExFT.
This was described by the little generalised metric \eqref{NClittlem}, which we reproduce here:
\be
m_{ab} = \begin{pmatrix} 
- e^{-4/5} \tau_\mu \tau_\nu & 0 & 0 &  e^{6/5} \varepsilon_{\mu\rho} \hat v^\rho \\
0 & e^{-4/5} & 0 & 0 \\
0 & 0 & e^{-4/5} & 0 \\
e^{6/5} \varepsilon_{\nu\rho} \hat v^\rho & 0 & 0 & - e^{6/5} 2 \tilde \Phi  
\end{pmatrix} \,.
\ee
We can attempt to relate this to a Riemannian parametrisation by U-dualising. 
Label the four-dimensional coordinates by $\hat i = ( \mu , u, 4)$.
Then we can U-dualise along both $\mu$ directions and $u$ using the transformation \eqref{UBuscher} with
\be
U^a{}_b = \begin{pmatrix} \delta^\mu_\nu & 0 & 0 & 0 \\ 0 & 1 & 0 & 0 \\ 0 & 0 & 0 & 1  \\ 0 & 0 & -1 & 0 \end{pmatrix} \,,
\ee
leading to a Riemannian parametrisation with 4-d metric
\be
g_{\hat i \hat j} = \begin{pmatrix} - e^{-4/3} \tau_\mu \tau_\nu & 0 & e^{2/3} \varepsilon_{\mu \rho} \hat v^\rho \\ 0 & e^{-4/3} & 0 \\ e^{2/3} \varepsilon_{\nu \rho} \hat v^\rho & 0  &- e^{2/3} 2 \tilde \Phi \end{pmatrix}\,.
\ee
Alternatively we can dualise along one of $\mu$ directions, $u$ and $4$ using 
\be
U^a{}_b = \begin{pmatrix} \delta^\mu_\nu - n^\mu n_\nu  & 0 & 0 & n^\mu \\ 0 & 1 & 0 & 0 \\ 0 & 0 & 1 & 0  \\ -n_\nu & 0 & 0 & 0 \end{pmatrix} \,.
\ee
The generalised metric becomes
\be
m_{ab} = \begin{pmatrix}
K_{\mu\nu} & 0 & 0 & c_\mu \\
0 & e^{-4/5} & 0 & 0 \\ 
0 & 0 & e^{-4/5} & 0 \\
c_\nu & 0 & 0 & e^{-4/5} ( n^\rho \tau_\rho )^2 
\end{pmatrix} \,,
\ee
where we let
\be
\begin{split}
K_{\mu\nu} & = e^{-4/5} ( \tau_\mu \tau_\nu - ( n^\rho \tau_\rho) ( \tau_\mu n_\nu + \tau_\nu n_\mu) + ( n^\rho \tau_\rho)^2 n_\mu n_\nu )
\\ & \qquad\qquad
 + e^{6/5} \left( - 2 n_{(\mu} \varepsilon_{\nu) \rho} \hat v^\rho + 2 n_\mu n_\nu n^\rho \varepsilon_{\rho \sigma} \hat v^\sigma  - 2 \tilde \Phi n_\mu n_\nu \right)
 \\ 
 c_\mu & = e^{-4/5} ( \tau_\mu n^\rho \tau_\rho - ( n^\rho \tau_\rho)^2 n_\mu ) - e^{6/5} n_\mu n^\rho \varepsilon_{\rho \sigma} \hat v^\sigma \,.
\end{split}
\ee
If we take $n^\mu = - \hv^\mu$, $n_\mu = \tau_\mu$ then we get a purely metric parametrisation with
\be
g_{\hat i \hat j} = e^{-4/3} \begin{pmatrix}
2 \tilde \Phi e^2 \left( \tau_\mu \tau_\nu + \frac{1}{2\tilde \Phi} \tau_{(\mu} \varepsilon_{\nu) \rho} \hat v^\rho \right) & 0 & 0 \\
0 & 1 & 0 \\
0 & 0 & 1
\end{pmatrix} \,.
\ee

\section{The usual cosets} 
\label{usualcosets} 

In this subsection, we demonstrate for the cases of $\Gsix$ and $\Gseven$ how to compute the quantity $r$ defined in \eqref{r} which appears in the trace \eqref{traceP} of the projector $P_{MN}{}^{MN}$, and which encodes the vital information about the coset in which the generalised metric is valued. We will not study these particular ExFTs elsewhere in this paper. However, we think that the form of the verification that $r=0$ may be instructive for future generalisations of non-Riemannian parametrisations to these groups.

For the $d=6$ $\Gsix$ ExFT, $R_1$ is the fundamental $\mathbf{27}$.
The Y-tensor is given by
\begin{align}
Y^{MN}{}_{PQ} = 10 d^{MNK} d_{PQK}
\end{align}
with $d^{MNK}$ and $d_{MNK}$ the symmetric cubic invariants of $\Gsix$. We use the $\operatorname{USp}(8)$ construction of \cite{Musaev:2014lna} in which the generalised metric is formed from the generalised vielbein $V_M{}^{ij}$ carrying an antisymmetric pair of eight-dimensional indices $i,j$, which transform in the fundamental of $\Hsix$. These indices are raised and lowered with the symplectic form $\Omega_{ij}$ such that
\begin{align}\label{eq:VOmega}
V_{Mij} = V_M{}^{kl} \Omega_{ki} \Omega_{lj}\,, \qquad V_M{}^{ij} \Omega_{ij} = 0\,,
\quad \Omega_{ik} \Omega^{jk} = \delta_i^j\,,
\end{align}
and then $\mathcal{M}_{MN} = V_M{}^{ij} V_{Nij}$.
The orthogonality relations with the inverse vielbein are
\begin{align}
V_M{}^{ij}V_{ij}{}^N = \delta_M^N, \qquad V_M{}^{kl}V_{ij}{}^M = \delta^{kl}_{ij} - \frac{1}{8} \Omega_{ij} \Omega^{kl}.
\end{align}
Finally, the totally symmetric invariant $d^{MNK}$ is given in terms of the symplectic form as
\begin{align}
d^{MNK} = \frac{2}{\sqrt{5}} V_{ij}{}^M V_{kl}{}^N V_{mn}{}^P \Omega^{jk} \Omega^{lm} \Omega^{ni}\,,\quad 
d_{MNK}  = \frac{2}{\sqrt{5}} V_M{}^{ij} V_N{}^{kl} V_P{}^{mn} \Omega_{jk} \Omega_{lm} \Omega_{ni}.
\end{align}
We can then calculate 
\be
\begin{split}
\mathcal{M}_{MN} d^{MNK} & =\frac{2}{\sqrt{5}} V_M{}^{pq} V_N{}^{rs}\Omega_{rp} \Omega_{sq} V_{ij}{}^M V_{kl}{}^N V_{mn}{}^P \Omega^{jk} \Omega^{lm} \Omega^{ni}\\ 
& = \frac{2}{\sqrt{5}} \left(\delta^{pq}_{ij} - \frac{1}{8} \Omega_{ij} \Omega^{pq} \right) \left(\delta^{rs}_{kl} - \frac{1}{8} \Omega_{kl} \Omega^{rs} \right) V_{mn}{}^P \Omega_{rp} \Omega_{sq} \Omega^{jk} \Omega^{lm} \Omega^{ni}\\
& = \frac{2}{\sqrt{5}} V_{mn}{}^P \left( \frac{1}{2} ( \Omega_{ki} \Omega_{lj} - \Omega_{kj} \Omega_{li} ) - \frac{1}{8} \Omega_{ij} \Omega_{kl} \right) \Omega^{jk} \Omega^{lm} \Omega^{ni}\\
& \propto V_{mn}{}^P \Omega^{mn}
\end{split}
\ee
which vanishes by \eqref{eq:VOmega}. 

For $\Gseven$, \cite{Hohm:2013uia}, $R_1$ is the fundamental $\mathbf{56}$.
Denote the generators acting on the fundamental by $t_{\alpha}{}_M{}^N$, and the antisymmetric invariant by $\Omega_{MN}$, with inverse $\Omega^{MN}$ such that $\Omega^{MK} \Omega_{NK} = \delta_N^M$.
The Y-tensor is then
\be
Y^{MN}{}_{KL} = - 12 t_{\alpha}{}^{MN} t^\alpha{}_{KL} - \frac{1}{2} \Omega^{MN} \Omega_{KL} \,,
\ee
where we defined $t_{\alpha MN} \equiv t_{\alpha M}{}^K \Omega_{KM}$, $t_\alpha{}^{MN} \equiv t_{\alpha K}{}^N \Omega^{MK}$ are both symmetric in $MN$. 
The adjoint projector is
\be
\mathbb{P}_{adj}{}^K{}_M{}^L{}_N = t_{\alpha M}{}^K t^\alpha{}_N{}^L = \frac{1}{24} \left( \delta_M^K\delta_N^L + 2 \delta_M^L \delta_N^K - \Omega_{MN} \Omega^{KL} \right) + t_{\alpha MN} t^{\alpha KL} \,,
\ee
such that $\mathbb{P}_{adj}{}^N{}_M{}^M{}_N = 133$ and hence $t_{\alpha MN} t^{\alpha MN} = -133$. Now, we introduce a generalised vielbein carrying antisymmetrised $\mathrm{SU}(8)$ indices, $E_M{}^{\bar M} = ( E_M{}^{AB}, E_{M AB} )$, such that \cite{Godazgar:2014nqa}
\be
\gM_{MN} = E_M{}^{\bar M} E_N{}^{\bar N} \bar \gM_{\bar M \bar N} \equiv  E_M{}^{AB} E_{N AB} + E_{N}{}^{AB} E_{M AB} \,.
\ee
Here $\bar\gM_{\bar M \bar N}$ represents the flat generalised metric.
As $\Omega^{MN} \gM_{MN} = 0$ by symmetry, we can show that $Y^{MN}{}_{KL} \gM_{MN} \gM^{KL}$ vanishes by proving that $t_{\alpha}{}^{MN} \gM_{MN} = 0$. 
We have
\be
t_{\alpha}{}^{MN} \gM_{MN} = 
t_{\alpha}{}^{MN} E_M{}^{\bar M} E_{N}{}^{\bar N} \bar\gM_{\bar M \bar N} 
= E^{\bar \alpha}{}_{\alpha} t_{\bar \alpha}{}^{\bar M \bar N} \bar \gM_{\bar M \bar N} \,,
\ee
where $E^{\bar \alpha}{}_{\alpha}$ is the adjoint representation of the vielbein (which we do not need) and $t_{\bar \alpha}{}^{\bar M \bar N}$ corresponds to the $\Gseven$ generator in the $\mathrm{SU}(8)$ basis. 
In this basis $V^{\bar M} = ( V^{AB }, V_{AB} )$ and $V_{\bar \alpha} = ( V_A{}^B, V_{ABCD} )$, $\Omega_{AB}{}^{CD} = \delta_{AB}^{CD}$ and (see e.g. appendix of \cite{LeDiffon:2011wt}) the components of $t_{\bar \alpha}{}^{\bar M \bar N}$ are then
\be
\begin{split} 
( t_A{}^B )_{CD}{}^{EF}& = - \delta^B_{[C} \delta^{E F}_{D]A} - \frac{1}{8} \delta_A^B \delta^{EF}_{CD} = + ( t_A{}^B )^{EF}{}_{CD}{}\,,\\
( t_{ABCD} )_{EFGH} &= \frac{1}{4!} \eta_{ABCDEFGH} \,,\quad ( t_{ABCD} )^{EFGH} = - \delta_{ABCD}^{EFGH}\,.
\end{split}
\ee
We then want to compute
\be
t_{\bar \alpha}{}^{\bar M \bar N} \bar \gM_{\bar M \bar N} = 2 t_{\bar \alpha}{}^{AB}{}{}_{CD} \bar{\gM}_{AB}{}^{CD}\,,
\ee
which is automatically zero for $\bar \alpha = {}_{ABCD}$ and for $\bar \alpha = {}_A{}^B$ turns out to vanish on evaluating the contractions.
We conclude that $t_{\alpha}{}^{MN} \gM_{MN} = 0$, hence $(P_{R_2}){}_{MN}{}^{PQ} \gM_{PQ} = 0$, hence $Y^{PQ}{}_{MN} \gM_{PQ} = 0$.

\bibliography{NewBib}

\end{document}